\documentclass[iop]{emulateapj}
\usepackage{ifpdf}
\ifpdf
\usepackage[iop]{emulateapj}
\else

\fi

\slugcomment{To appear in AJ}

\shorttitle{Molecular Hydrogen in NGC 4565 and NGC 5907}
\shortauthors{Laine et al.}

\begin{document}

\newcommand{\HI}{H\,{\sc i}}
\newcommand{\HII}{H\,{\sc ii}}
\newcommand{\Ht}{${\rm H_{2}}$}
\newcommand{\mic}{$\mu$m}

\title{Warm Molecular Hydrogen Emission in Normal Edge-On Galaxies \\NGC 4565 and NGC 5907}

\author{Seppo Laine}
\affil{{\it Spitzer} Science Center - Caltech, MS 220-6,
Pasadena, CA 91125; seppo@ipac.caltech.edu}

\author{Philip N. Appleton}
\affil{NASA Herschel Science Center, California Institute of Technology, Pasadena, CA 91125, USA; apple@ipac.caltech.edu}

\author{Stephen T. Gottesman}
\affil{Department of Astronomy, University of Florida, Gainesville,
FL 32611-2055; gott@astro.ufl.edu}

\author{Matthew L. N. Ashby}
\affil{Harvard-Smithsonian Center for Astrophysics, 60 Garden Street, Cambridge, MA 02138;
mashby@cfa.harvard.edu}

\and

\author{Catherine A. Garland}
\affil{Natural Sciences Department, Castleton State College, Castleton, VT 05735; catherine.garland@castleton.edu}

\begin{abstract} 

We have observed warm molecular hydrogen in two nearby edge-on disk galaxies,
NGC~4565 and NGC~5907, using the {\it Spitzer} high-resolution infrared
spectrograph. The 0--0 S(0) 28.2~$\mu$m and 0--0 S(1) 17.0~$\mu$m pure rotational
lines were detected out to 10 kpc from the center of each galaxy on both sides
of the major axis, and in NGC~4565 the S(0) line was detected at $r$ = 15 kpc on
one side. This location is beyond the transition zone where diffuse neutral
atomic hydrogen starts to dominate over cold molecular gas, and marks a
transition from a disk dominated by high surface-brightness far-IR emission to
that of a more quiescent disk. It also lies beyond a steep drop in the radio
continuum emission from cosmic rays in the disk. Despite indications that star
formation activity decreases with radius, the \Ht\ excitation temperature and
the ratio of the \Ht\ line and the far-IR luminosity surface densities,
$\Sigma$(L$_{H2}$)/$\Sigma$(L$_{TIR}$), change very little as a function of
radius, even into the diffuse outer region of the disk of NGC~4565. This
suggests that the source of excitation of the \Ht\ operates over a large range
of radii, and is broadly independent of the strength and relative location of UV
emission from young stars. Although excitation in photodissociation regions is
the most common explanation for the widespread \Ht\ emission, cosmic ray heating
or shocks cannot be ruled out. At $r$ = 15~kpc in NGC~4565, outside the main UV
and radio continuum-dominated disk, we derived a higher than normal \Ht\ to
7.7~$\mu$m PAH emission ratio, but this is likely due to a transition from
mainly ionized PAH molecules in the inner disk to mainly neutral PAH molecules
in the outer disk. The inferred mass surface densities of warm molecular
hydrogen in both edge-on galaxies differ substantially, being 4(--60) M$_{\sun}$
pc$^{-2}$ and 3(--50) M$_{\sun}$~pc$^{-2}$ at $r$ = 10 kpc for NGC~4565 and
NGC~5907, respectively. The higher values represent very unlikely point-source upper limits. The point source case is not supported by the observed emission distribution in the spectral slits. These mass surface densities cannot support the observed rotation velocities in excess of 200~km~s$^{-1}$. Therefore, warm molecular hydrogen cannot account for dark matter in these disk  galaxies, contrary to what was implied by a previous {\it ISO} study of  the nearby edge-on galaxy NGC~891.

\end{abstract}


\section{INTRODUCTION}

\begin{figure*}
\centering
\includegraphics[width=15cm]{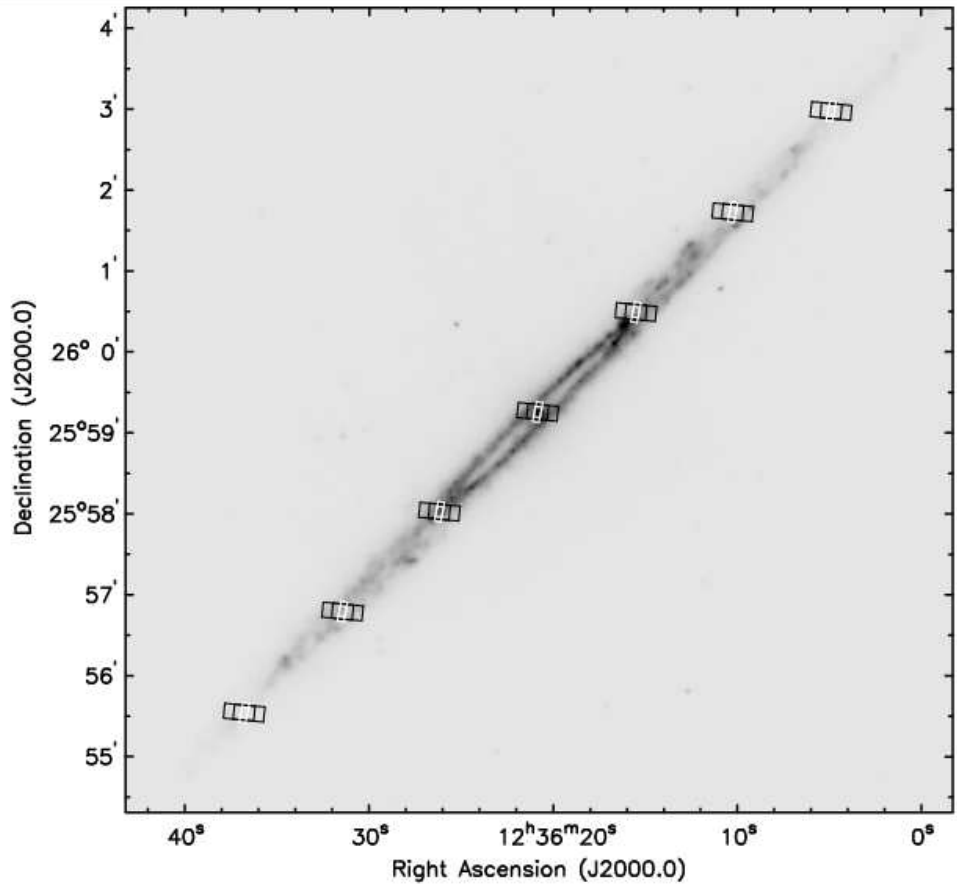}
\caption{Positions of the IRS SH (white) and LH (black) slits overlaid on an
8~\mic\ {\it Spitzer}/IRAC image of NGC~4565. Spectra from $\sim$~10 to 37~\mic\
were taken at the nucleus and at distances of 5, 10, and 15 kpc from the
nucleus  along the galaxy's major axis. Emission from warm
\Ht\ was detected even at the farthest northwestern (top right) position.}
\label{fig1}
\end{figure*}

\begin{figure}
\centering
\includegraphics[width=10cm]{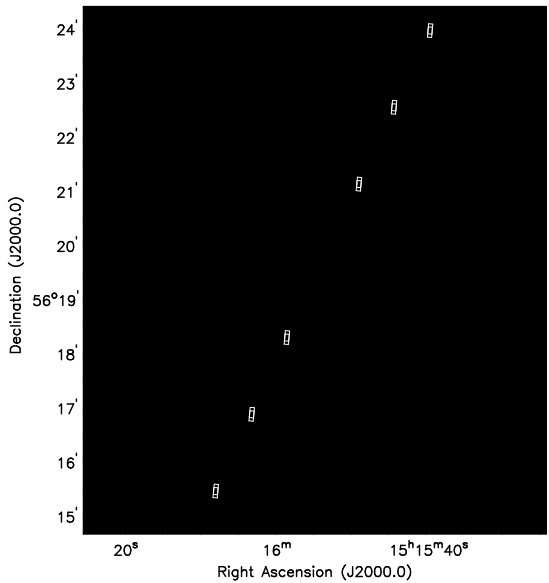}
\caption{Positions of the IRS SH (white) and LH (black) slits overlaid on an
8~\mic\ {\it Spitzer}/IRAC image of NGC~5907. Spectra from $\sim$~10 to 37~\mic\
were taken at the nucleus and at distances of 5, 10, and 15 kpc from the
nucleus along the galaxy's major axis.}
\label{fig2}
\end{figure}

The physical conditions and excitation mechanisms of atomic and molecular gas in
the outer disks of nearby spiral galaxies are only beginning to be explored.  Most
of this gas is thought to be neutral atomic hydrogen (\ion{H}{1}),  or cold (T $<$
50 K) molecular hydrogen (\Ht). The presence of cold \Ht\ is usually inferred only
by indirect means via observations of carbon monoxide (CO), and quantified by
assuming an (uncertain) empirical conversion factor between the two molecules.
Direct detection of \Ht\ is preferable. However, since \Ht\ has no allowed dipole
radiative transitions, it has to be heated above $\sim$~100~K to radiate
significantly via quadrupole pure-rotational transitions in the mid-infrared
(mid-IR) or through ro-vibrational transitions from even warmer gas emerging in the
near-infrared. Furthermore, the other direct observational window -- the detection
through the absorption of UV radiation in the electronic Lyman--Werner bands -- is
challenging, and only under rare conditions has it been possible to detect the
presence of cold \Ht\ in the Galaxy through FUV absorption
\citep[e.g.,][]{snow2000,rachford2009}.

\begin{figure*}
\centering
\includegraphics[width=10cm]{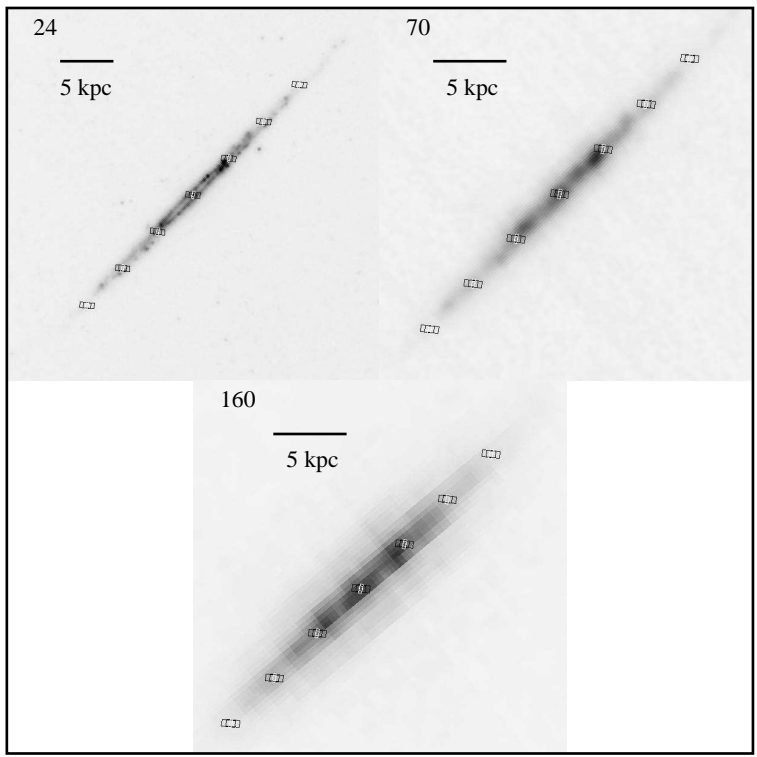}
\caption{Archival {\it Spitzer}/MIPS images of NGC 4565 at 24~\mic\ (upper left), 70~\mic~(upper right), and 160~\mic\ (bottom). The SH (white) and LH (black) slit positions have been overlaid.}
\label{fig3}
\end{figure*}

The {\it Infrared Space Observatory (ISO)} provided the first opportunity to directly
observe warm extragalactic molecular hydrogen in nearby galaxies, unhampered by the
atmosphere \citep[e.g.,][]{rigo96,vale96,vale99}. More recently the  {\it Spitzer
Space  Telescope} \citep{werner04} has provided a wealth of new data on rotational
\Ht\ emission lines in dozens of nearby galaxies, ranging from normal galaxies
\citep[e.g.,][]{roussel07,brunner08} to Ultraluminous Infrared Galaxies
\citep[ULIRGs;][]{armus06}. Unusually strong intergroup \Ht\ emission associated with
a large-scale ($\sim$~30~kpc) X-ray emitting shock has recently been found associated
with the compact Stephan's Quintet galaxy group \citep[][]{apple06,cluver10} and the
{\it Taffy Galaxy} bridge (B. W. Peterson et al. 2010, in preparation). Similarly large \Ht\ line fluxes have
been found in 17 galaxies in a sample of 55 low-luminosity radio galaxies
\citep[][]{ogle2007,ogle2010}. In Stephan's Quintet and in the low-luminosity radio
galaxies, very weak thermal continua are detected, suggesting shock excitation of \Ht,
rather than excitation via photodissociation regions (PDRs) associated with star
formation \citep[e.g.,][]{guillard09b}. Other sources of \Ht\ heating, for example
cosmic ray heating, have also been suggested to explain the strong \Ht\ emission in
the Orion bar \citep{shaw09}.

It is therefore of interest to examine the strength of \Ht\ emission in the outer
regions of nearby galaxies, where star formation and cosmic ray heating are much
reduced, and where the dominant gas component is usually assumed to be neutral atomic
hydrogen rather than molecular gas. By outer regions in this paper we  mean the radii
10 kpc and beyond from the nucleus. {\it ISO} observations of the nearby edge-on
galaxy NGC~891 directly detected the abundant warm \Ht\ out to a distance of 11~kpc
from the galaxy center \citep{vale99}, with warm \Ht\ mass surface densities of
$\sim$~3000~M$_{\sun}$~pc$^{-2}$. This suggested a dominant contribution of molecular
hydrogen to the mass-density of the disk, and perhaps that molecular hydrogen could
contribute a significant part of the ``missing mass'' in this galaxy.

Intrigued by these early {\it ISO} results, we pursued {\it Spitzer} Infrared
Spectrograph \citep[IRS;][]{houck04} observations of two local, nearly edge-on
galaxies, NGC~4565 and NGC~5907, to explore the possibility of massive reservoirs
of warm molecular gas far from the nuclei. These early-mission IRS high resolution
spectra cover infrared wavelengths from 10~\mic\ to 37~\mic, and target the 0--0
S(0) and 0--0 S(1) \Ht\ lines, which are known to contain the strongest emission
from the mass in warm molecular gas in nearby galaxies \citep[e.g.,][]{roussel07}.
The spectral range also covered several other mid-IR lines which assisted us in
exploring the importance of star formation as an excitation mechanism in these
regions. Although the current observations are not extremely sensitive, they
provide interesting constraints on the nature of \Ht\ emission in the outer disks
of galaxies. 

\begin{figure*}
\centering
\includegraphics[width=15cm]{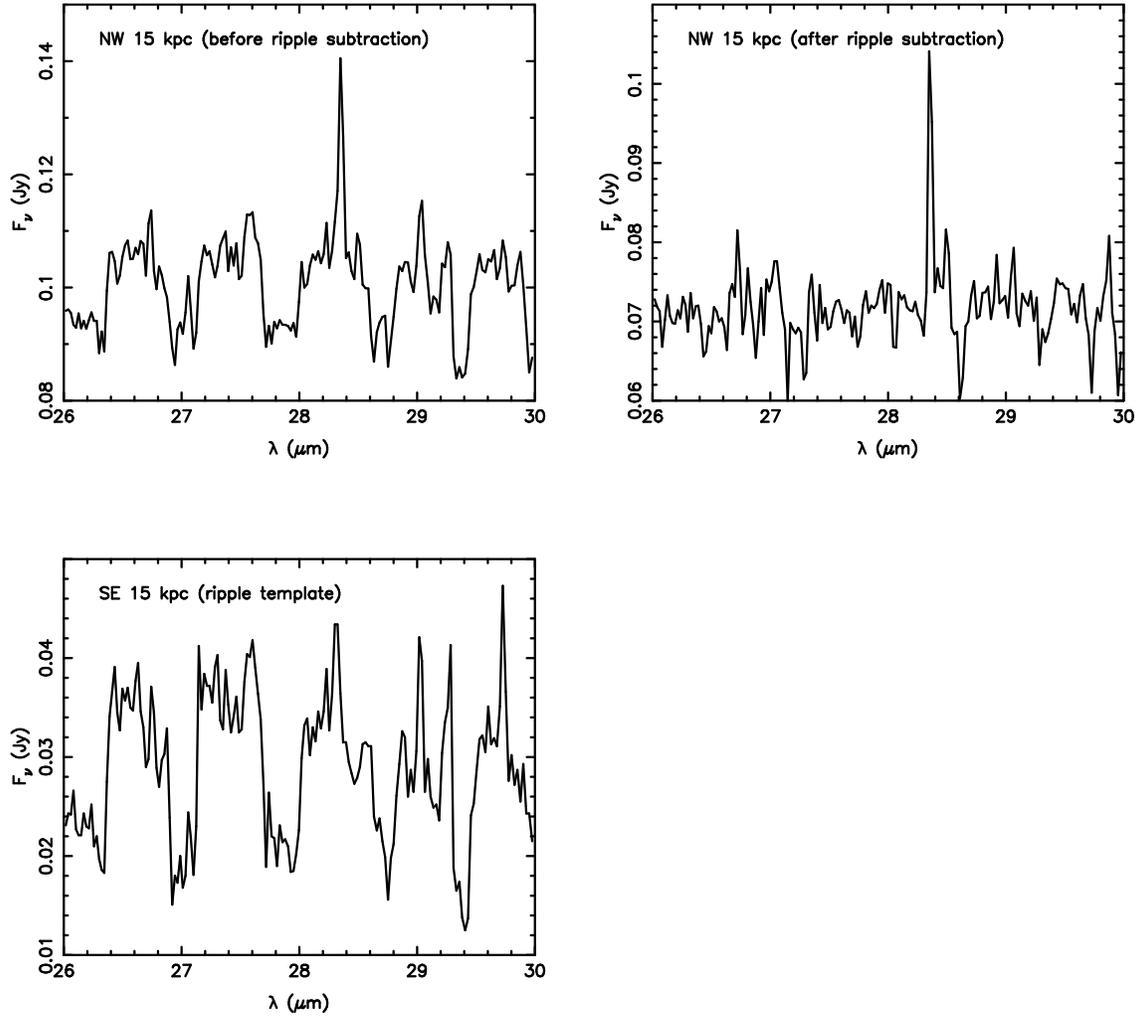}
\caption{Ripple spectrum subtraction. The original spectrum around the 28 micron \Ht\ line in NGC 4565 at 15 kpc NW is at top left, the same spectrum after the ``ripple spectrum'' subtraction is at top right, and the ``ripple spectrum'' from the SE 
15~kpc position is at bottom left.}
\label{fig4}
\end{figure*}

To assist in our analysis, we also utilized {\it Spitzer} Infrared Array Camera
(IRAC) 8~\mic\ images of NGC~4565 (Figure \ref{fig1}) and NGC~5907 (Figure
\ref{fig2}) taken in {\it Spitzer} program PID 3 (P.I. Giovanni Fazio; M. L. N.
Ashby 2009, private communication). Finally, we utilized archival {\it Spitzer}
Multiband Imaging Photometer (MIPS) images of NGC~4565 and NGC~5907 at 24, 70, and
160~\mic\ (Figure~\ref{fig3} shows  these maps for NGC~4565).

\begin{figure*}
\centering
\epsscale{0.7}
\includegraphics[width=13cm]{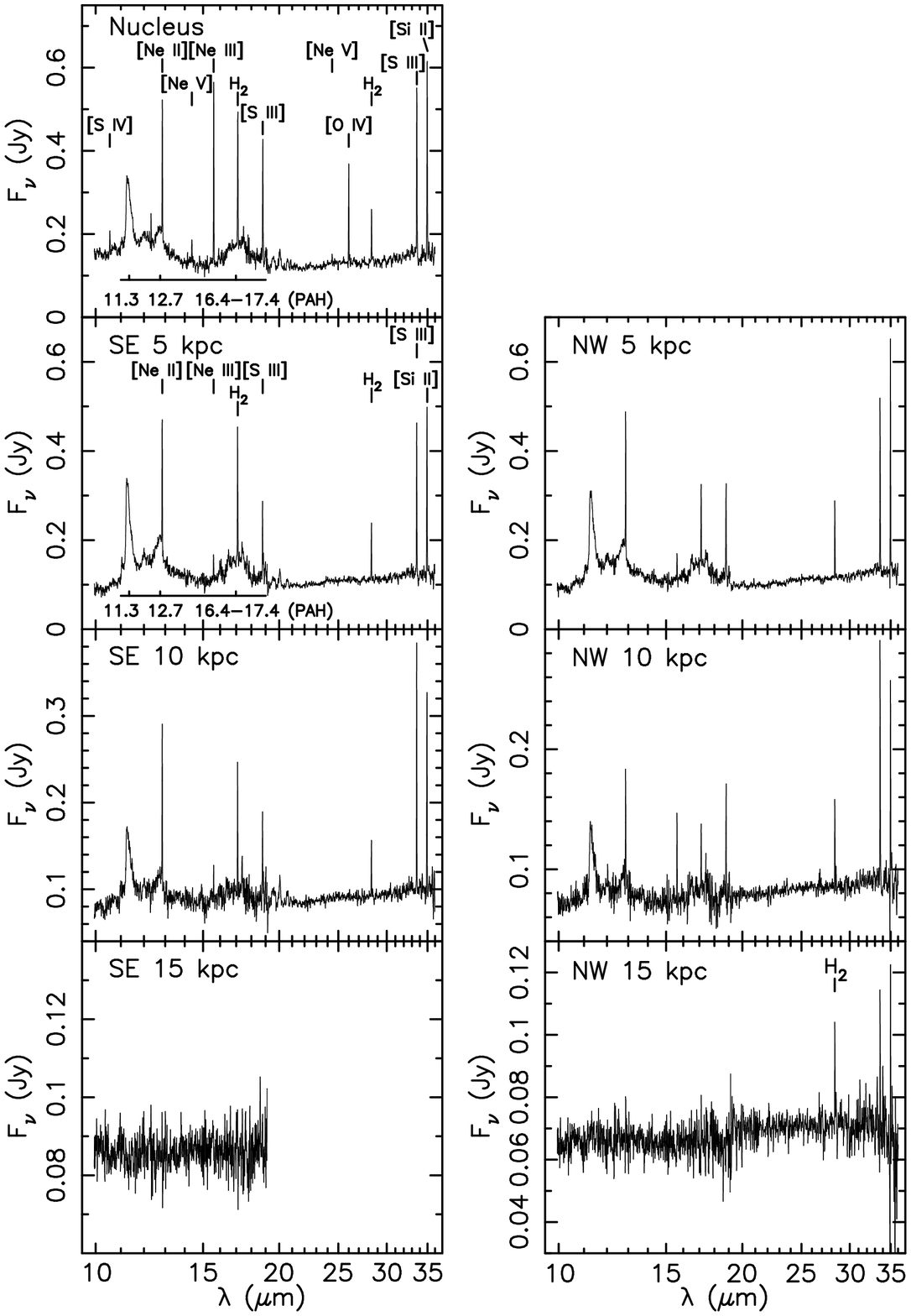}
\caption{Combined SH and LH spectra of NGC~4565 at the seven positions shown in Figure \ref{fig1}. The observed flux density in Jy is plotted versus the observed wavelength in \mic. \Ht\ is detected at all locations except the 15~kpc~SE pointing. A variety of forbidden lines and broad PAH regions are also identified.}
\label{fig5}
\end{figure*}

NGC~4565 is a nearby (we adopted a distance of 10 Mpc for our observations), Sb-type
nearly edge-on (inclination 88\degr; \citeauthor{alton04} \citeyear{alton04}) large
(D$_{25}$=15\farcm 8) disk galaxy with a nucleus classified as Sy1.9 \citep{deo07}.
A sharp dust lane delineates the disk plane of the galaxy, and there is significant
obscuration caused by dust within the galactic plane. \citet{nein96} found that this
galaxy has a nuclear molecular disk as well as a molecular gas ring at a distance of
$\sim$~1\arcmin--2\arcmin\ (3--6 kpc) from the nucleus, and weaker extended
molecular gas emission. The molecular gas ring has an associated dust ring, which is
seen in the {\it Spitzer} 8~\mic~ image shown in Figure \ref{fig1}. The \ion{H}{1}
distribution is asymmetric along the disk plane, with substantially more emission
coming from the northwestern side, and there is a strong, continuous warp in the
\ion{H}{1} emission starting at $\sim$~7\arcmin\ on both sides of the nucleus
\citep{rupen91}. \citet{sofue94a} showed that at a radius of $\sim$~10~kpc the
interstellar medium (ISM) transitions from being dominated by molecular gas to
being dominated by atomic gas.

\begin{figure*}
\centering
\includegraphics[width=13cm]{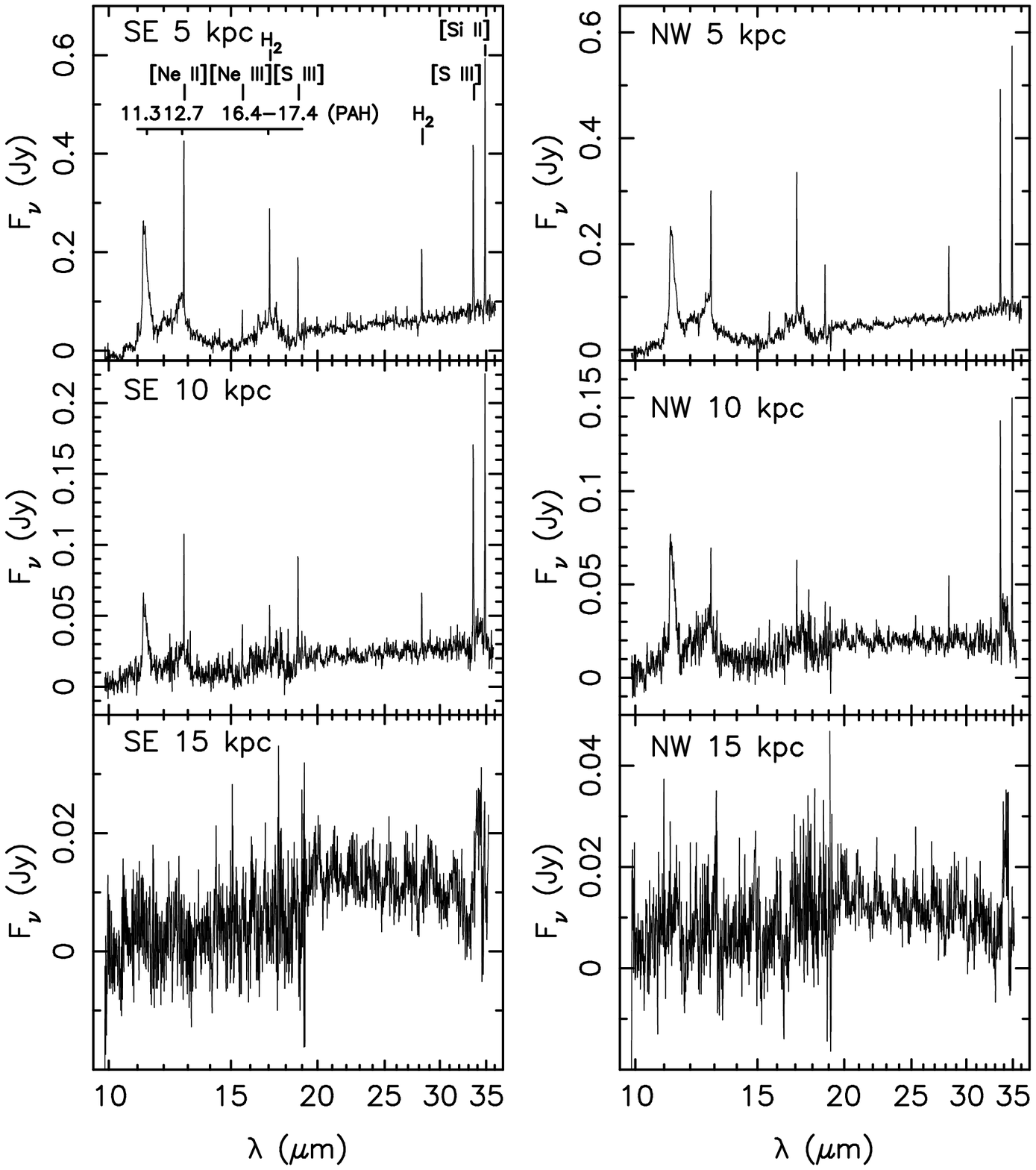}
\caption{Combined SH and LH spectra of NGC~5907 at the six positions 
shown in Figure \ref{fig2}. The observed flux density in Jy is plotted versus the observed wavelength in \mic. \Ht\ is detected at all but the two outermost
pointings. A variety of forbidden lines and broad PAH regions are also
identified.}
\label{fig6}
\end{figure*}

NGC~5907 is a similarly large (D$_{25}$=12\farcm 6), nearby (adopted distance 11
Mpc), almost edge-on (inclination 87\degr; \citeauthor{alton04} \citeyear{alton04}),
disk galaxy. CO observations show a fast-rotating nuclear molecular disk with
bar-like non-circular motions beyond the nucleus \citep*{garcia97}. The \ion{H}{1}
distribution shows a warp at both the southeastern and northwestern sides of the
nucleus, starting at $\sim$~5\arcmin\ radius \citep{sancisi76,shang98}.

\begin{table*}[!ht]
\caption{Forbidden lines in NGC~4565.\label{table1a}}
\begin{tabular}{cccccc}
\tableline\tableline
\\[0.25pt]
Position & [NeII]12.81$\mu$m\tablenotemark{a} & [NeIII]15.55$\mu$m\tablenotemark{a}  & [SIII]18.71$\mu$m\tablenotemark{a} & [SIII]33.48$\mu$m\tablenotemark{b} & [SiII]34.81$\mu$m\tablenotemark{b}\\
\tableline
\\[0.25pt]
SE 5 kpc  Ex  & 187$\pm$6        & 44$\pm$5            & 65$\pm$6            & 335$\pm$13         & 397$\pm$15  \\
SE 5 kpc  Pt  & 283$\pm$9        & 73$\pm$9            & 118$\pm$10          & 531$\pm$21         & 650$\pm$15  \\
SE 10 kpc Ex  & 95$\pm$5         & 28$\pm$12         & 48$\pm$4            & 284$\pm$6          & 234$\pm$13  \\
SE 10 kpc Pt  & 144$\pm$7        & 47$\pm$20           & 88$\pm$8            & 386$\pm$9          & 383$\pm$13  \\
SE 15 kpc Ex  & 30$\pm$7         & $<$ 14              & 13$\pm$3             & \nodata            & \nodata     \\
SE 15 kpc Pt  & 45$\pm$11        & $<$ 23              & 24$\pm$5             & \nodata            & \nodata     \\
NW 5 kpc  Ex  & 207$\pm$6        & 40$\pm$5            & 87$\pm$3           & 411$\pm$8         & 487$\pm$7  \\  
NW 5 kpc  Pt  & 314$\pm$9        & 67$\pm$9            & 158$\pm$6           & 652$\pm$13         & 799$\pm$11  \\  
NW 10 kpc Ex  & 69$\pm$3        & 41$\pm$5            & 36$\pm$3            & 199$\pm$6         & 144$\pm$5   \\ 
NW 10 kpc Pt  & 104$\pm$5        & 68$\pm$7            & 65$\pm$5            & 316$\pm$10         & 236$\pm$8   \\ 
NW 15 kpc Ex  & 28$\pm$8        & 28$\pm$8           & $<$ 10              & 40$\pm$5           & 69$\pm$8  \\
NW 15 kpc Pt  & 43$\pm$12        & 47$\pm$13           & $<$ 19              & 64$\pm$8           & 113$\pm$13  \\
\end{tabular}
\tablecomments{Measured line fluxes are given in 10$^{-19}$ W~m$^{-2}$.}
\end{table*}

\begin{deluxetable*}{cccccccccc}
\tabletypesize{\scriptsize}
\tablecaption{Forbidden lines in the nucleus of NGC~4565. \label{table1b}}
\tablewidth{0pt}
\tablehead{
\colhead{} & \colhead{SIV~10.51\tablenotemark{a}} & \colhead{NeII~12.81\tablenotemark{a}} & \colhead{NeV~14.32\tablenotemark{a}} & \colhead{NeIII~15.55\tablenotemark{a}} & \colhead{SIII~18.71\tablenotemark{a}} & \colhead{NeV~24.31\tablenotemark{b}} & \colhead{OIV~25.89\tablenotemark{b}} & \colhead{SIII~33.48\tablenotemark{b}} & \colhead{SiII~34.82\tablenotemark{b}}
}
\startdata
Pt & 55$\pm$9 & 322$\pm$6 & 47$\pm$3 & 354$\pm$8 & 238$\pm$13 & 52$\pm$7 & 431$\pm$11 & 726$\pm$13 & 867$\pm$21 \\
Ex & 39$\pm$6 & 213$\pm$4 & 30$\pm$2 & 212$\pm$5 & 131$\pm$7 & 37$\pm$5 & 302$\pm$8 & 457$\pm$8 & 529$\pm$13 \\
\enddata
\tablenotetext{a}{Integrated over the SH aperture (53 arcsec$^2$).}
\tablenotetext{b}{Integrated over the LH aperture (248 arcsec$^2$).}
\tablecomments{Measured line fluxes are given in 10$^{-19}$ W~m$^{-2}$. The numbers after the ion species give the rest wavelength in microns.}

\end{deluxetable*}

\begin{deluxetable}{lccc}
\tabletypesize{\scriptsize}
\tablecaption{\Ht\ line fluxes in NGC~4565. \label{table1c}}
\tablewidth{0pt}
\tablehead{\colhead{Position} & \colhead{0--0~S(0)\tablenotemark{b}} & \colhead{0--0~S(1)\tablenotemark{a}} & \colhead{0--0~S(2)\tablenotemark{a}}  \\
}
\startdata
Nucleus Ex  & 189$\pm$5     & 170$\pm$7        & 29$\pm$6 \\   
Nucleus Pt  & 281$\pm$8     & 298$\pm$14       & 43$\pm$10 \\   
SE 5 kpc Ex & 176$\pm$3  &121$\pm$4    & $<$ 12 \\
SE 5 kpc Pt & 261$\pm$5  & 213$\pm$7   & $<$ 17     \\
SE 10 kpc Ex &  88$\pm$1   & 52 $\pm$2     & $<$ 17 \\    
SE 10 kpc Pt & 130$\pm$2   & 92$\pm$4      & $<$ 25 \\    
SE 15 kpc Ex &  \nodata    & $<$ 14        & $<$ 16 \\
SE 15 kpc Pt & \nodata     &  $<$ 25       & $<$ 23 \\
NW 5 kpc  Ex & 221$\pm$4   & 83$\pm$3      & $<$ 13 \\   
NW 5 kpc  Pt & 327$\pm$6   & 145$\pm$5     & $<$ 19 \\   
NW 10 kpc Ex & 89$\pm$5    & 36$\pm$3      & $<$ 19 \\
NW 10 kpc Pt & 132$\pm$7   &  64$\pm$5     & $<$ 27 \\
NW 15 kpc Ex &  36$\pm$2   & $<$ 13        & $<$ 13 \\
NW 15 kpc Pt & 53$\pm$3    & $<$ 22        & $<$ 18 \\
\enddata
\tablenotetext{a}{Integrated over the SH aperture (53 arcsec$^2$).}
\tablenotetext{b}{Integrated over the LH aperture (248 arcsec$^2$).}
\tablecomments{The measured line fluxes are given in 10$^{-19}$ W~m$^{-2}$.}

\end{deluxetable}

\section{OBSERVATIONS}
\label{observ}

We observed NGC~4565 and NGC~5907 with both high resolution modules of {\it
Spitzer's} IRS instrument on 2005 January 10 and 2005 June 6, respectively (PID
3319; \dataset{ADS/Sa.Spitzer\#00010745344} and
\dataset{ADS/Sa.Spitzer\#00010745088}).  The short-high (SH) module covers
wavelengths from 9.9 to 19.6~\mic~and has a slit size of 4\farcs 7 $\times$
11\farcs 3, while the long-high (LH) module brackets  the 18.7 to
37.2~\mic~wavelength range with a slit size of 11\farcs 1 $\times$ 22\farcs 3.
We took one cycle of ``staring mode'' observations with a 120 s ramp time with
the SH module and a 240 s ramp time with the LH module. The effective
integration times were approximately doubled to $\sim$~240~s (SH) and
$\sim$~480~s (LH) because each cycle takes two spectra, moving the target to
positions 1/3 and 2/3 slit lengths away from the end of the slit along the slit
long axis. We observed three positions along the galaxy major axes on both sides
of the nuclei at distances of  5, 10, and 15 kpc from the nucleus. We also
observed the nucleus of NGC~4565. Projections of the SH and LH slits on the
8~\mic~IRAC galaxy images are shown in Figures \ref{fig1} and \ref{fig2}. By
overlaying our observed positions on visible light and \ion{H}{1} maps we 
confirmed that the gaseous and stellar warps start beyond the outermost
observed locations in these two galaxies. Only in the northwest 15 kpc pointing
in NGC~5907 could a very small amount of \Ht\ have been missed if it strictly
follows the \ion{H}{1} distribution. However, even in that position the
majority of \ion{H}{1} emission comes from along the major axis of the
galaxy.

\begin{deluxetable*}{cccccc}
\tabletypesize{\scriptsize}
\tablecaption{Forbidden line fluxes in NGC~5907. \label{table2a}}
\tablewidth{0pt}
\tablehead{\colhead{Position} & \colhead{[NeII]12.81$\mu$m\tablenotemark{a}} & \colhead{[NeIII]15.55$\mu$m\tablenotemark{a}}  & \colhead{[SIII]18.71$\mu$m\tablenotemark{a}} & \colhead{[SIII]33.48$\mu$m\tablenotemark{b}} & \colhead{[SiII]34.81$\mu$m\tablenotemark{b}}\\
}
\startdata
SE 5 kpc Ex & 187$\pm$5           & 33$\pm$2            & 70$\pm$4          & 386$\pm$21        & 553$\pm$9 \\
SE 5 kpc Pt & 284$\pm$7           & 55$\pm$4            & 127$\pm$7          & 613$\pm$33        & 906$\pm$14 \\

SE 10 kpc Ex & 48$\pm$2           & 21$\pm$3            & 30$\pm$2            & 197$\pm$14        & 201$\pm$7 \\
SE 10 kpc Pt & 72$\pm$3           & 35$\pm$5            & 54$\pm$4           & 312$\pm$22        &  329$\pm$11 \\
 
SE 15 kpc Ex & $<$ 12             & $<$ 13              & $<$ 13             & $<$ 21            & $<$ 31 \\
SE 15 kpc Pt & $<$ 18             & $<$ 22              & $<$ 24             & $<$ 34            & $<$ 51 \\

NW 5 kpc Ex & 168$\pm$8          & 31$\pm$5            & 50$\pm$2           & 379$\pm$12        & 481$\pm$4 \\
NW 5 kpc Pt & 254$\pm$12          & 52$\pm$8            & 90$\pm$4           & 602$\pm$19        & 789$\pm$6 \\

NW 10 kpc Ex & 28$\pm$2           & 8$\pm$0.4          & 13$\pm$2           & 115$\pm$10        & 146$\pm$4 \\
NW 10 kpc Pt & 43$\pm$3           & 14$\pm$0.6          & 24$\pm$3           & 182$\pm$16        & 239$\pm$7 \\

NW 15 kpc Ex & $<$ 14             & $<$ 7              & $<$ 18             & $<$ 22            &$<$ 32 \\
NW 15 kpc Pt & $<$ 21             & $<$ 12              & $<$ 33             & $<$ 35            &$<$ 53 \\
\enddata
\tablenotetext{a}{Integrated over the SH aperture (53 arcsec$^2$).}
\tablenotetext{b}{Integrated over the LH aperture (248 arcsec$^2$).}
\tablecomments{The measured line fluxes are in 10$^{-19}$ W~m$^{-2}$.}
\end{deluxetable*}

The background brightnesses (due to ecliptic emission) were $\sim$~29 MJy
sr$^{-1}$ for NGC~4565, and 17--18 MJy sr$^{-1}$ for NGC~5907. No separate
background spectra were taken, since the recommended observing strategy during
the first cycle of {\it Spitzer} observations was still evolving and no clear
recommendations existed at the time. This considerably complicated the removal
of bad pixels from the spectra, discussed below. We did not use the ``peak-up''
option since our targets are extended, but the intrinsic IRS pointing accuracy
of $\sim$~1$^\prime$$^\prime$ was sufficient for our purposes.

We used spectra that were processed through the standard {\it Spitzer} IRS
pipeline (version S13.2.0). We first edited the basic calibrated data frames to
remove bad or ``rogue'' pixels, using a custom-made software script that allows
interactive removal of isolated bad pixels from these data. We then ran the
spectra through the S17 version of the custom spectral extraction software SPICE
provided by the {\it Spitzer} Science Center, using the whole slit width
extractions and initially the standard point source calibration for flux
calibration.

Corrections were later made to line fluxes for both point and extended source
calibration using the slit-loss factors provided in SPICE, and these two extreme
limits bracket the true (unknown) distribution of gas in the slits. A comparison
of the \Ht\ 28~$\mu$m line fluxes for the two nod positions in both galaxies
shows differences of $<$10\%, suggesting the gas distribution is relatively
smooth on the size scale of the nods (five and nine arcseconds, respectively, 
for SH and LH slits). Extended emission is also suggested by 
structure along the slit in individual images. We consider the
distribution of the emission-line gas to be close to flat and extended, and apply
the corresponding calibration, but also quote, for reference, the very unlikely 
values derived by applying the point source calibration, when interpreting the 
properties of the observed galaxies.

For the LH spectra of NGC~4565, we encountered a low-level ``fringing'' effect
not seen in the spectra of NGC~5907. This effect would normally have been removed
had we obtained dedicated ``off'' observations (not obtained during our Cycle-1
observations), and appears to be the result of incomplete ``jail bar'' removal in
the pipeline (an effect in which parts of the detector array show a patterning,
which in this case was brighter than usual).  To remove this effect, we decided
to use the observations taken at 15 kpc southeast (SE) of the nucleus of NGC~4565
as a reference, and subtracted this LH spectrum from all the others. This led to
a significant improvement in the spectra. Nonetheless, such a procedure runs the
risk that there may have been faint emission at that reference position which
would be removed from all other LH points (this primarily affects the \Ht\ 0--0
S(0) line which lies in the LH module. However, unlike the 15 kpc northwest (NW)
point, which clearly shows 28~$\mu$m S(0) emission even before we performed the
subtraction, the SE point appears  devoid of emission as can be seen in
Figure~\ref{fig4}. We feel confident, therefore, that the use of the SE 15 kpc
spectrum as a reference has not adversely affected our conclusions. Although
faint emission might have been present in this reference spectrum, based on
measurements of the raw spectrum of NGC~4565 at the 15 kpc SE position, we
believe that an rms upper limit to such emission is 1.2 $\times$ 10$^{-21}$
W~m$^{-2}$~Hz$^{-1}$ at the position of the 0--0 S(0) line, corresponding to less
than 10$\%$ of the faintest \Ht\ emission detected at the 15 kpc NW point. In
other words, the subtraction of the reference spectrum from the observations
introduces a systematic error (not to be confused with a random error) which is
estimated to be less than 10\% of the faintest emission detected, a result that
does not affect the conclusions of this paper. None of the SH spectra were
affected by this instrumental effect.

Finally, we combined (by averaging) the spectra obtained at the two nod positions at
each separate radial distance. The final spectra are shown in Figures~\ref{fig5} and
\ref{fig6}.

\section{RESULTS}

The 0--0 S(0) and 0--0 S(1) transitions of \Ht\ were detected at the 5 and 10~kpc
distances from the nucleus in both galaxies. \Ht\ was also detected at the northwest
15~kpc location in NGC~4565. On this side of the galaxy there is also substantially 
more \ion{H}{1} emission \citep{rupen91}. The 0--0 S(2) transition of
\Ht\ was also detected in the Seyfert nucleus of NGC~4565. A variety of forbidden
lines ([\ion{Ne}{2}]~12.81~\mic, [\ion{Ne}{3}]~15.55~\mic,
[\ion{S}{3}]~18.71/33.48~\mic, and [\ion{Si}{2}]~34.82~\mic) were detected in most
locations of both galaxies.  

The continuum emission from the nucleus of NGC~4565 appears relatively flat,
although it shows the broad PAH emission feature around 17~\mic. Because the IRS
apertures cover several hundred parsecs, most of this PAH emission is likely emitted
by the disk of this galaxy. In addition to the forbidden lines detected elsewhere in
this galaxy, [\ion{S}{4}]~10.51~\mic, [\ion{Ne}{5}]~14.32/24.31~\mic, and
[\ion{O}{4}]~25.89~\mic\ were detected  in the nucleus. The indicator lines of
active galactic nuclei (AGNs), [\ion{Ne}{5}]~14.32~\mic\ and
[\ion{Ne}{5}]~24.31~\mic\ \citep[e.g.,][]{armus04}, were both detected at a
signal-to-noise ratio of $\sim$10.  

\subsection{Line Fluxes}

We extracted line fluxes by fitting Gaussians to the lines using the SMART software
package \citep{higdon04}. We fitted the broad aromatic features (indicated in
Figures \ref{fig5} and \ref{fig6}) with Lorentzian profiles \citep[cf. the
Lorentzian method used by][]{galliano08}. The extracted fluxes are given in
Tables~\ref{table1a}, \ref{table1b}, \ref{table1c}, \ref{table2a}, and
\ref{table2b}, where ``Pt'' indicates point source calibrated spectra and ``Ex''
indicates fluxes corresponding to a flat, infinitely extended distribution, as
explained in detail in Section 2.

The uncertainties were estimated by taking into account the quality of the
profile fit. Generally high signal-to-noise ratios ($>$~10) were obtained for
most of the lines except in the outer regions of the disks. Upper limits were
estimated as 4 $\times~\sigma~\times~\delta\lambda$, where $\sigma$ is the rms
noise in the region of the expected line and $\delta$$\lambda$ is the width of
an unresolved line at the corresponding wavelength (which corresponds essentially to the width of the bandpass for the high resolution modules, $\lambda$/600).

To assist in diagnosing the gas excitation conditions, we also estimated the flux
densities in the IRAC 8~\mic\ images of NGC~4565 and NGC~5907, under the areas
covered by the IRS slits in our observations. We used the same IRAC 8~\mic\ filter
width as \citet{roussel07} to convert the flux densities from Jy into fluxes in
W~m$^{-2}$, but we have not attempted to subtract the stellar emission from the IRAC
image as it is generally only a few per cent of the total emission at 8~\mic.  

\begin{deluxetable}{cccc}
\tabletypesize{\scriptsize}
\tablecaption{\Ht\ line fluxes in NGC~5907. \label{table2b}}
\tablewidth{0pt}
\tablehead{\colhead{Position}  & \colhead{0--0 S(0)\tablenotemark{b}}     & 
\colhead{0--0 S(1)\tablenotemark{a}} & \colhead{0--0 S(2)\tablenotemark{a}} \\
}
\startdata
SE 5 kpc Ex & 180$\pm$7  & 92$\pm$5   &$<$ 16 \\
SE 5 kpc Pt & 267$\pm$10  & 162$\pm$8   &$<$ 24 \\

SE 10 kpc Ex & 51$\pm$2    & 18$\pm$2    &$<$ 12  \\
SE 10 kpc Pt & 75$\pm$3    & 32$\pm$3    &$<$ 18  \\

SE 15 kpc Ex & $<$ 17      & $<$ 9       &$<$ 9  \\
SE 15 kpc Pt & $<$ 25      & $<$ 16      &$<$ 14  \\

NW 5 kpc Ex & 188$\pm$3   & 104$\pm$3   &$<$ 13  \\
NW 5 kpc Pt & 278$\pm$4   & 182$\pm$6   &$<$ 20  \\
 
NW 10 kpc Ex & 51$\pm$3   & 15$\pm$2     &$<$ 15  \\
NW 10 kpc Pt & 76$\pm$4   & 27$\pm$3     &$<$ 22  \\

NW 15 kpc Ex & $<$ 18      & $<$ 13      &$<$ 16  \\
NW 15 kpc Pt & $<$ 26      & $<$ 23      &$<$ 24  \\
\enddata
\tablenotetext{a}{Integrated over the SH aperture (53 arcsec$^2$).}
\tablenotetext{b}{Integrated over the LH aperture (248 arcsec$^2$).}
\tablecomments{The measured line fluxes are in 10$^{-19}$ W~m$^{-2}$.}
\end{deluxetable}

\subsection{Ionized Gas Line and PAH Feature Ratios}

\begin{figure*}
\centering
\includegraphics[width=10cm,angle=270]{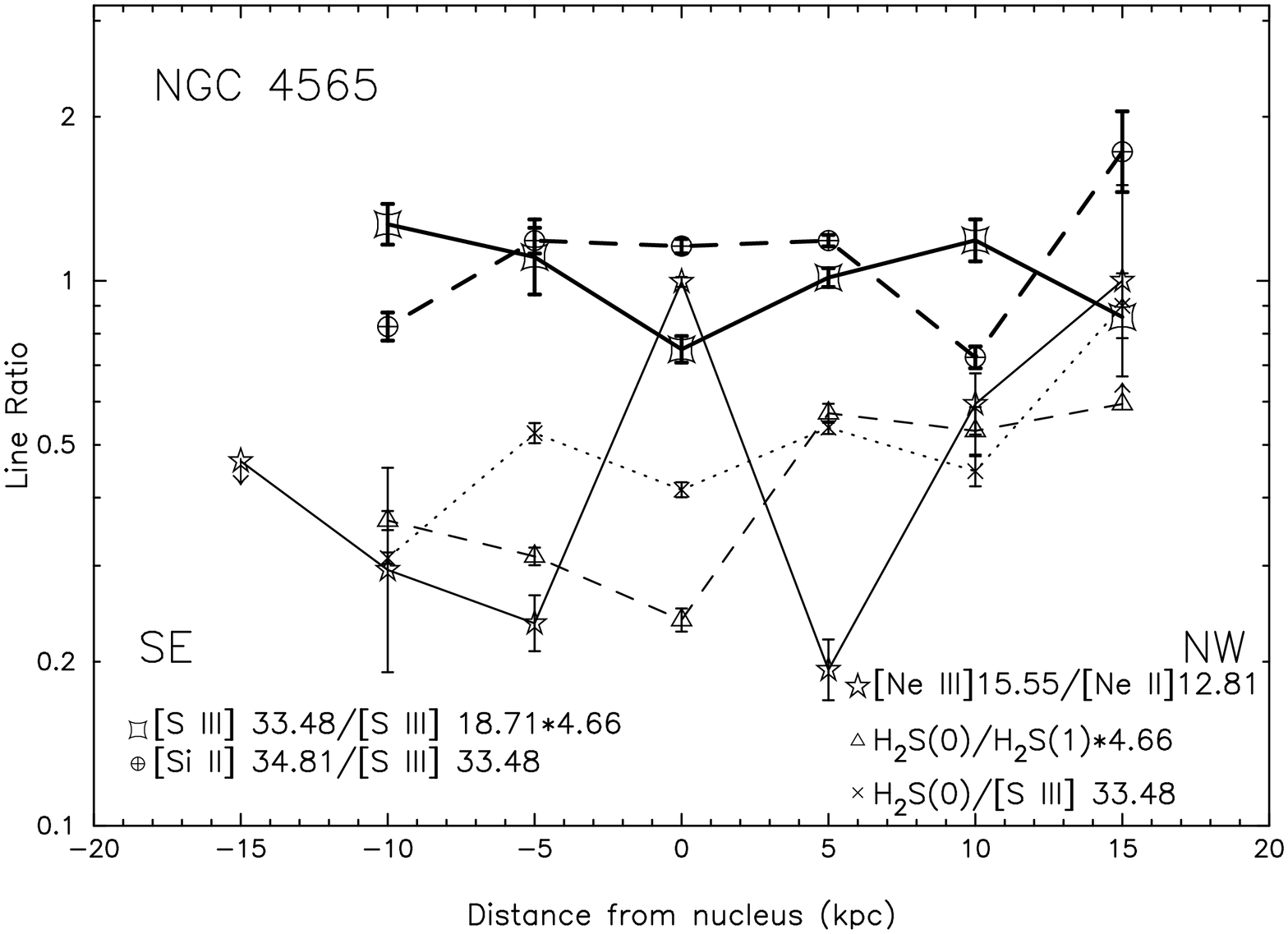}
\caption{Line ratios in NGC~4565. The solid thin line is for [\ion{Ne}{3}]/[\ion{Ne}{2}], the dashed thin line is for \Ht\ S(0)/\Ht\ S(1),
the dotted line is for \Ht\ S(0)/[\ion{S}{3}] 33.48, the thick solid line is for
[\ion{S}{3}] 33.48/[\ion{S}{3}] 18.71, and the dashed thick line is for [\ion{Si}{2}] 34.81/[\ion{S}{3}] 33.48.}
\label{fig7}
\end{figure*}

\begin{figure*}
\centering
\includegraphics[width=10cm,angle=270]{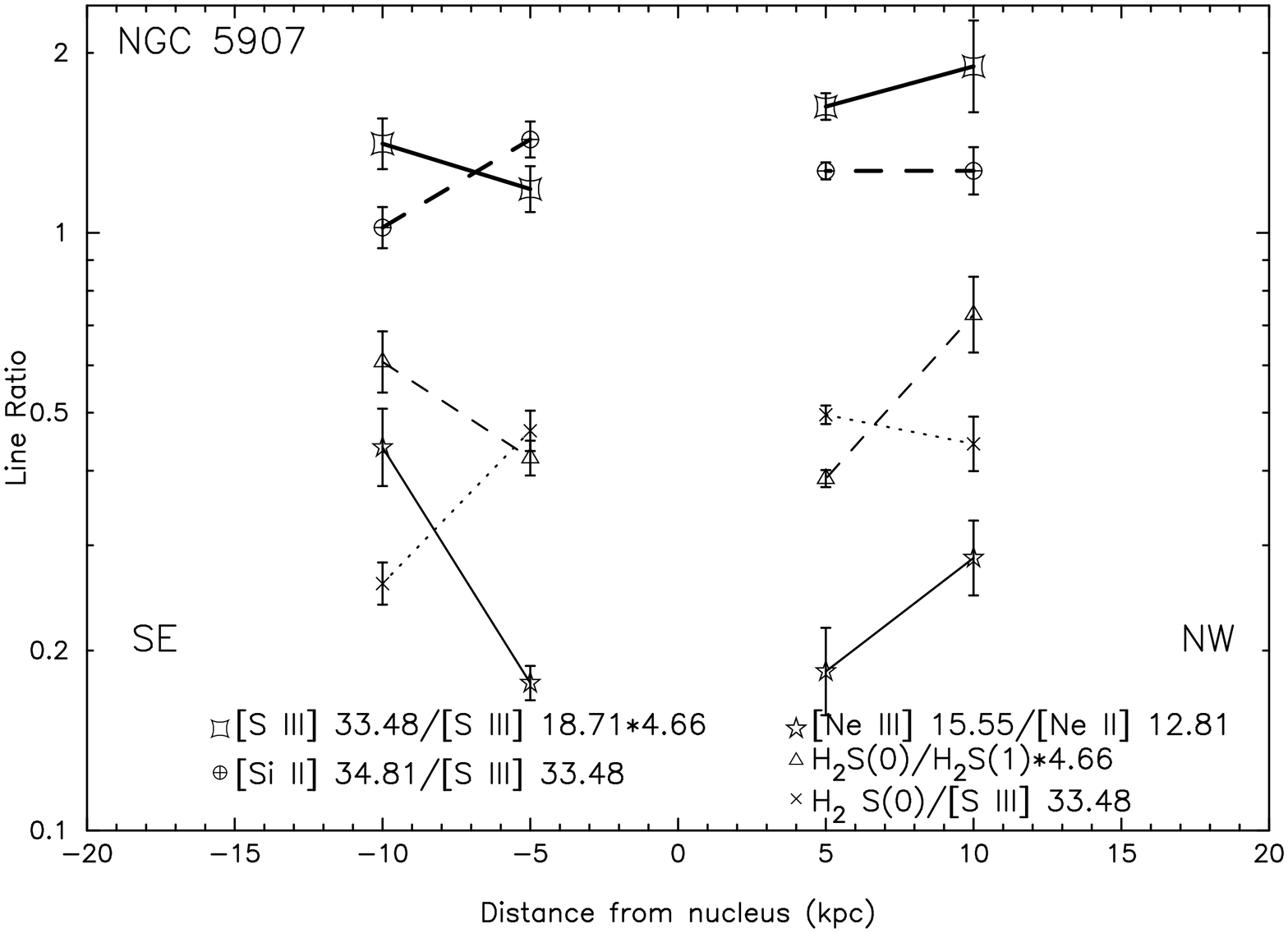}
\caption{Line ratios in NGC~5907. The solid thin line is for [\ion{Ne}{3}]/[\ion{Ne}{2}], the dashed thin line is for \Ht\ S(0)/\Ht\ S(1),
the dotted line is for \Ht\ S(0)/[\ion{S}{3}] 33.48, the thick solid line is for
[\ion{S}{3}] 33.48/[\ion{S}{3}] 18.71, and the dashed thick line is for [\ion{Si}{2}] 34.81/[\ion{S}{3}] 33.48.}
\label{fig8}
\end{figure*}

\begin{figure*}
\centering
\includegraphics[width=10cm,angle=270]{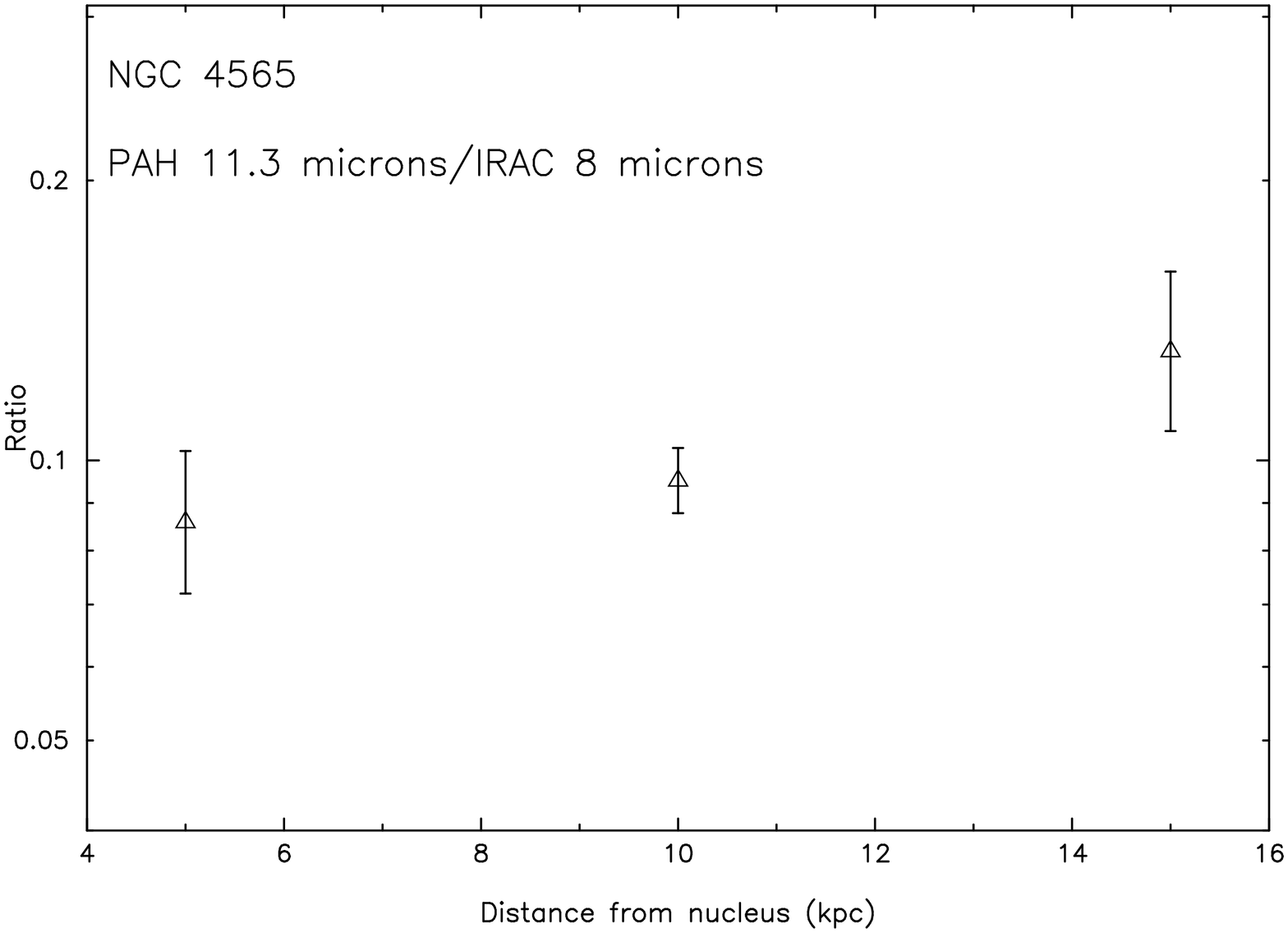}
\caption{Ratios of the fluxes in the 11.3~\mic\ and 7.7~\mic\ PAH features are plotted versus the distance from the nucleus (in kpc) on the northwestern side of NGC~4565. As explained in Section 3.2, the IRAC 8~\mic~flux is used  as a proxy for the 7.7~\mic\ PAH feature.}
\label{fig9}
\end{figure*}

To investigate the gas excitation conditions we compared the strengths of various
emission lines by forming line ratios. When comparing line ratios formed from lines
taken with two different modules (LH and SH), we scaled the fluxes by a factor of
4.66, the ratio of the areas of the two module apertures. We also applied the
extended source calibration correction to the line fluxes before taking the ratio
to be consistent with this approach.

\begin{figure*}
\centering
\epsscale{0.8}
\includegraphics[width=15cm]{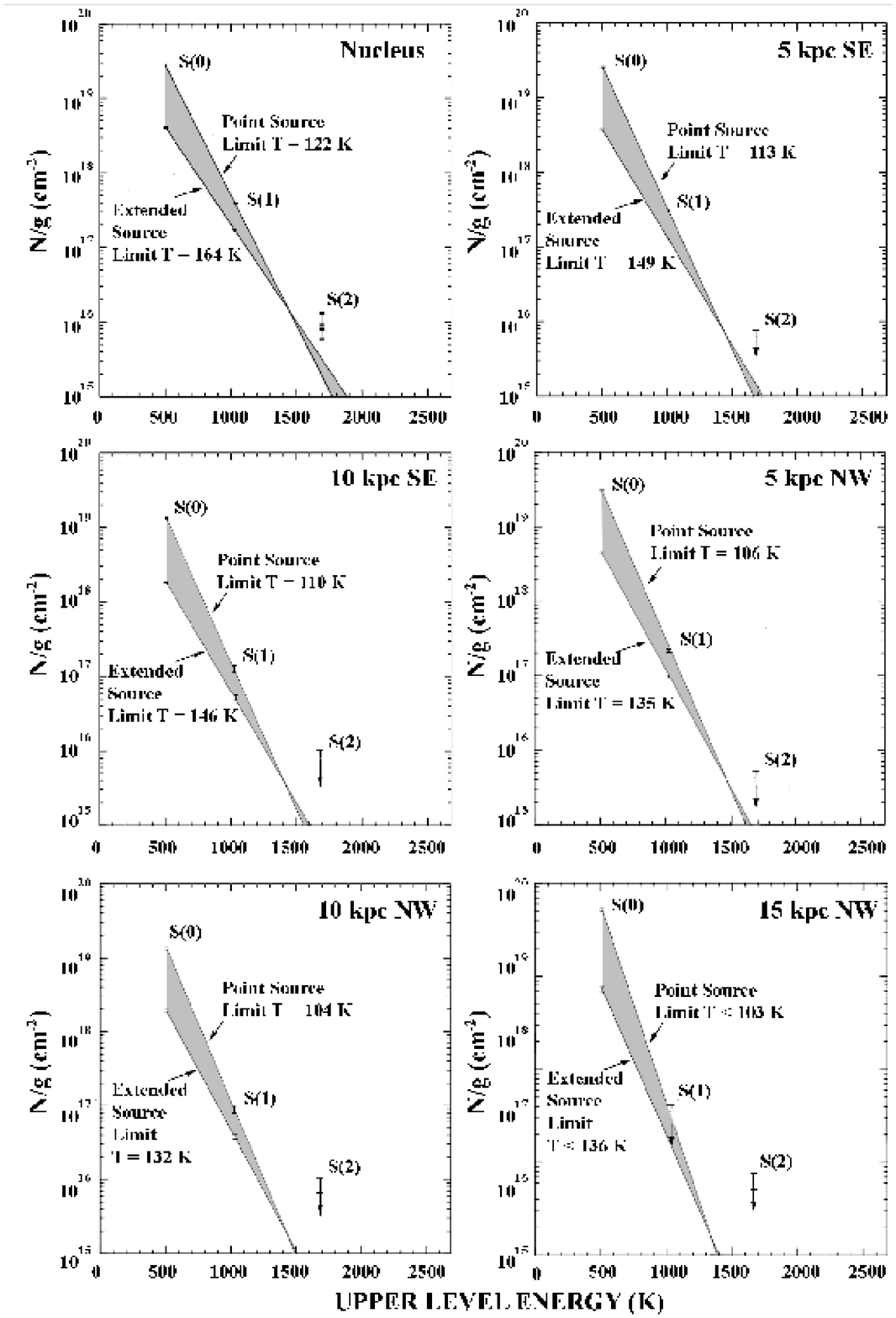}
\caption{Excitation diagrams for the disk positions of NGC~4565. The \Ht\ column density in the upper
level of the transition normalized by its statistical weight (in cm$^{-2}$) is
plotted versus the upper level energy (in K) for the S(0), S(1), and S(2)
transitions at all locations except 15 kpc~SE from the nucleus. Most S(2) data
points are only upper limits. The solid lines indicate the best fits to the
data points assuming a single-temperature component. The molecular hydrogen
gas temperatures implied by the fits are shown. As discussed in Section 3.3,
most likely there are multiple temperature components and the shown temperatures
are very likely upper limits.}
\label{fig10}
\end{figure*}

We show the line ratios in Figures~\ref{fig7} and \ref{fig8}. In NGC~4565 the
[\ion{S}{3}]~33.48~$\mu$m/[\ion{S}{3}]~18.71~$\mu$m ratios are close to 1,
typical for extranuclear regions seen in the SINGS sample of nearby galaxies
\citep{dale09}, except at 10 kpc SE where the ratio drops below 0.4. This would
imply a drop in the  electron density by factors of a few hundreds
\citep{dale09}. This ratio is slightly higher, between 1 and 2, in NGC~5907,
covering very well the region  in which most of the extranuclear areas studied by
\citet{dale09} fall. The [\ion{Si}{2}]~34.81~$\mu$m/[\ion{S}{3}]~33.48~$\mu$m
ratio in NGC~4565 has a  surprisingly low value of just above 1 at the nucleus,
which is at the lower end of values seen in the nuclei of AGN galaxies in the
sample of \citet{dale09}.

\begin{figure*}
\centering
\includegraphics[width=15cm]{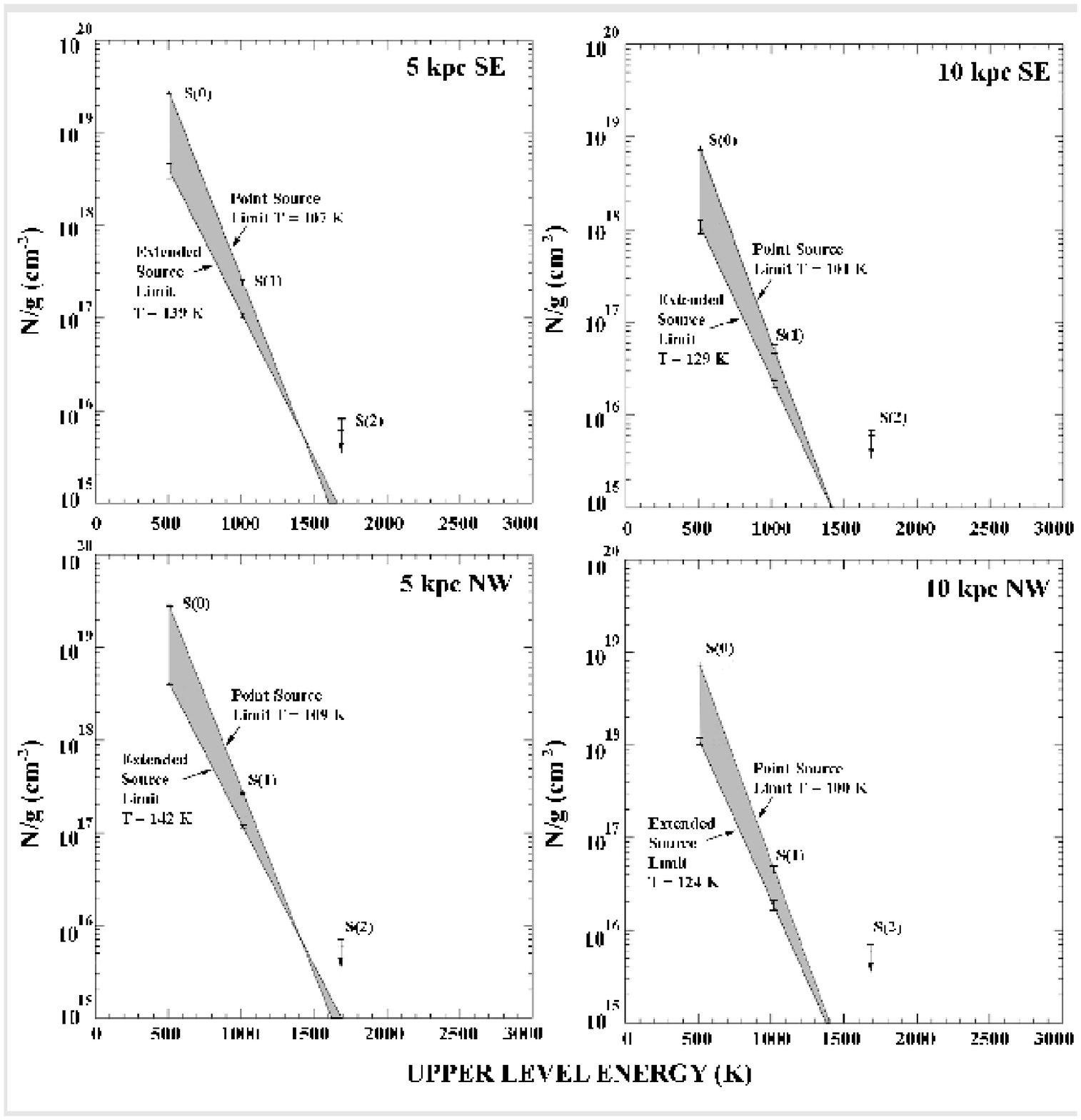}
\caption{Excitation diagrams for NGC~5907. The \Ht\ column density in the upper
level of the transition normalized by its statistical weight (cm$^{-2}$) is
plotted versus the upper level energy (K) for the S(0), S(1), and S(2)
transitions for the four inner locations. The S(2) data points are only upper
limits. The solid lines indicate the best fits to the data points assuming a
single-temperature component. The molecular hydrogen gas temperatures implied
by the fits are shown.}
\label{fig11}
\end{figure*}

\begin{deluxetable*}{lcccc}
\tablecaption{Physical parameters derived from the excitation diagrams for NGC~4565 \label{Excipars}}
\tablehead{
\colhead{Position} & \colhead{Temperature\tablenotemark{a}} & \colhead{Equilibrium o/p} &
\colhead{{N}$_{\rm H_2}$} & \colhead{${\Sigma}_{\rm H_2}$} \\
\colhead{(kpc)} & \colhead{(K)} & \colhead{} & \colhead{(10$^{20}$~mol~cm$^{-2}$)} & 
\colhead{(M$_{\sun}$~pc$^{-2}$)} \\
}
\startdata
Nuc & 164 -- (122) & 2.6 -- (2.0) & 4.1 -- (59) & 6.6 -- (95) \\
SE 5 & 149 -- (112) & 2.5 -- (1.9)  & 4.8 -- (74)  & 7.7 -- (119)\\
SE 10 & 146 -- (110) & 2.4 -- (1.8) & 2.4 -- (40) & 3.8 -- (64)\\
NW 5 & 135 -- (106) & 2.3 -- (1.7) & 7.5 -- (113) & 12. -- (182)\\
NW 10 & 132 -- (104) & 2.3 -- (1.7) & 3.3 -- (51) & 5.2 -- (82) \\
NW 15 & $<$136 -- (103) & 2.3 -- (1.6) & 1.1 -- (21) & 1.8 -- (34) \\
\enddata
\tablenotetext{a}{The values in parentheses are for the point source approximation.}
\end{deluxetable*}

\begin{deluxetable*}{lcccc}
\tablecaption{Physical parameters derived from the excitation diagrams for NGC~5907.\label{Excipars2}}
\tablehead{
\colhead{Position} & \colhead{Temperature\tablenotemark{a}} & \colhead{Equilibrium o/p} &
\colhead{{N}$_{\rm H_2}$} & \colhead{${\Sigma}_{\rm H_2}$} \\
\colhead{(kpc)} & \colhead{(K)} & \colhead{} & \colhead{(10$^{20}$~mol~cm$^{-2}$)} & 
\colhead{(M$_{\sun}$~pc$^{-2}$)} \\
}
\startdata
SE 5 & 139 -- (107) & 2.4 -- (1.8) & 5.8 -- (90)  & 9.3 -- (145)\\
SE 10 & 129 -- (101) & 2.2 -- (1.6) & 2.0 -- (32) & 3.3 -- (51)\\
NW 5 & 142 -- (109) & 2.4 -- (1.8) & 5.7 -- (89) & 9.2 -- (144)\\
NW 10 & 124 -- (100) & 2.1 -- (1.6) & 2.3 -- (33) & 3.6 -- (53)\\ 
\enddata
\tablenotetext{a}{The values in parentheses are for the point source approximation.}
\end{deluxetable*}

The [\ion{Ne}{3}]~15.55~\mic/[\ion{Ne}{2}]~12.81~\mic\ ratio behaves as expected
in both galaxies. It is higher in low-metallicity regions (towards larger radii
in both galaxies), and lower towards the center in regions that are expected to
have a higher metallicity. It achieves a high value in the nucleus of NGC~4565,
consistent with what was seen by \citet{dale09} in the SINGS sample, presumably
due to the higher excitation conditions near an AGN.

The \Ht\ S(0) to \Ht\ S(1) ratio hovers around 0.5 in both galaxies and is seen to
increase towards the outer disk in both galaxies on both sides of the disk. This may
primarily be an effect of the temperature, and it will be discussed in more detail
in Section~\ref{excisection}. We also show the \Ht\ S(0) to
[\ion{S}{3}]~33.48~$\mu$m ratio. [\ion{S}{3}]~33.48~$\mu$m is mostly excited by star
formation, and thus this ratio can be used as a rough indicator of the significance
of star formation induced excitation of the \Ht\ molecule. We see that the ratio
stays fairly constant at around 0.5 in both galaxies, but goes up at the 15 kpc NW
point in NGC~4565. This is consistent with the ionization level of molecules
dropping in the outermost disk, as discussed in Section~\ref{excitdiscussion} below.
Figure \ref{fig9} shows the ratios of the fluxes in the 11.3 and 7.7~\mic\ PAH
features versus the distance from the nucleus on the NW side of NGC~4565. The IRAC
8~\mic\ fluxes measured in the IRAC image within the SH aperture and at the same
spatial locations as the spectra were used as a proxy for the 7.7~\mic\ PAH flux.
The ratio is increasing towards 15 kpc NW, which most likely implies that the ISM is
becoming less ionized towards the outer disk, as the 7.7~\mic\ PAH feature consists
of more ionized dust material than the 11.3~\mic\ PAH feature
\citep*[e.g.,][]{alla99}.

\subsection{Excitation Diagrams}
\label{excisection}

\begin{figure*}
\centering
\includegraphics[width=10cm,angle=270]{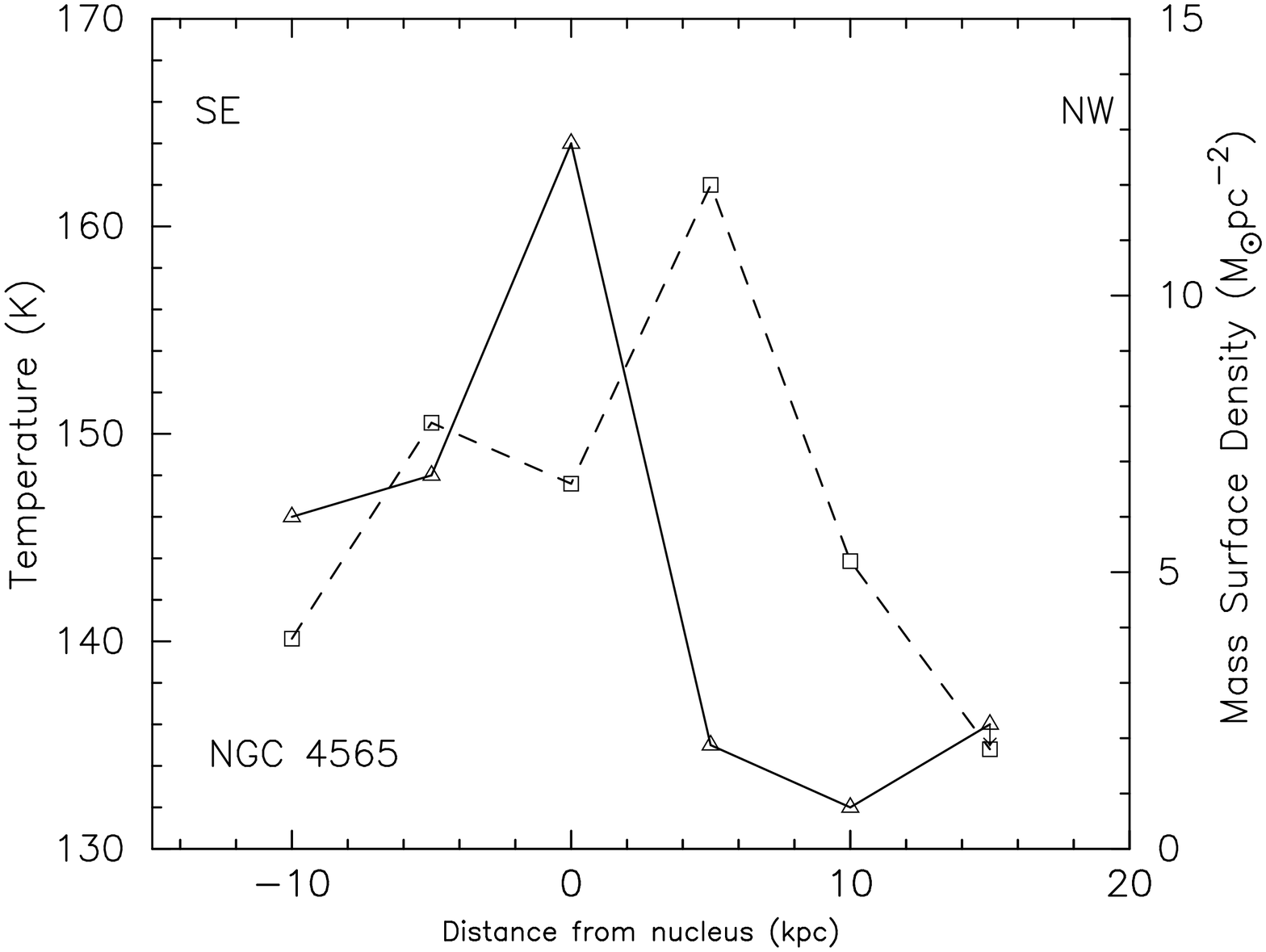}
\caption{Temperature in K (solid line) and the mass surface density in
M$_{\sun}$~pc$^{-2}$ (dashed line) are plotted versus distance from the
nucleus (in kpc) for NGC~4565. The estimates of temperature and mass surface density were derived from the \Ht\ S(0) and S(1) lines, and also the S(2) line in the nucleus. The temperature and mass surface density calculations were
made with the assumption that the warm molecular \Ht\ gas is uniformly 
distributed across the slits (see Sections~\ref{observ} and \ref{excisection} for discussion
on the flux distribution in the slits and the calibration differences between
smooth, extended and point source fluxes).}
\label{fig12}
\end{figure*}

We constructed excitation diagrams (Figures~\ref{fig10} and \ref{fig11}) from the \Ht\
data in order to place constraints on the molecular gas properties. These diagrams plot
the column density (N$_{\rm u}$) of \Ht\ in the upper level of each transition,
normalized by its statistical weight, versus the upper level energy E$_{\rm u}$
\citep[e.g.,][]{rigo02}, which we derived from the measured fluxes assuming local
thermodynamic equilibrium  for each position observed. Both extended and
point source flux distributions are shown for the detected lines.  The extended source
results include a wavelength-dependent slit-loss correction which makes the extended
flux calibration differ from the point source flux calibration typically by a factor of
~1.5, and has a further geometrical correction of 4.66 for the different areas of the
SH and LH slits. The grey area between the two limiting cases (point and extended) is
the parameter space that most likely encompasses the actual case. However, we stress
that the point source assumption is very unrealistic, as discussed in
Section~\ref{observ} and given the small slit aperture and the thickness of the disks.
As we will see, this assumption also leads to unlikely low gas temperatures close to
100 K, and therefore high implied \Ht\ gas surface densities. For this reason we prefer
to consider the extended source limit to be much closer to the actual situation, but
the point source provides a useful (although unrealistic) boundary. The uncertainty in
the \Ht\ properties is governed largely by this uncertainty in the slit loss
corrections rather  than the formal errors, which are quite small because the lines
were all detected with  quite high signal to noise ratios (SNR; they vary from 10 to 50
in most cases).

\begin{figure*}
\centering
\includegraphics[width=10cm,angle=270]{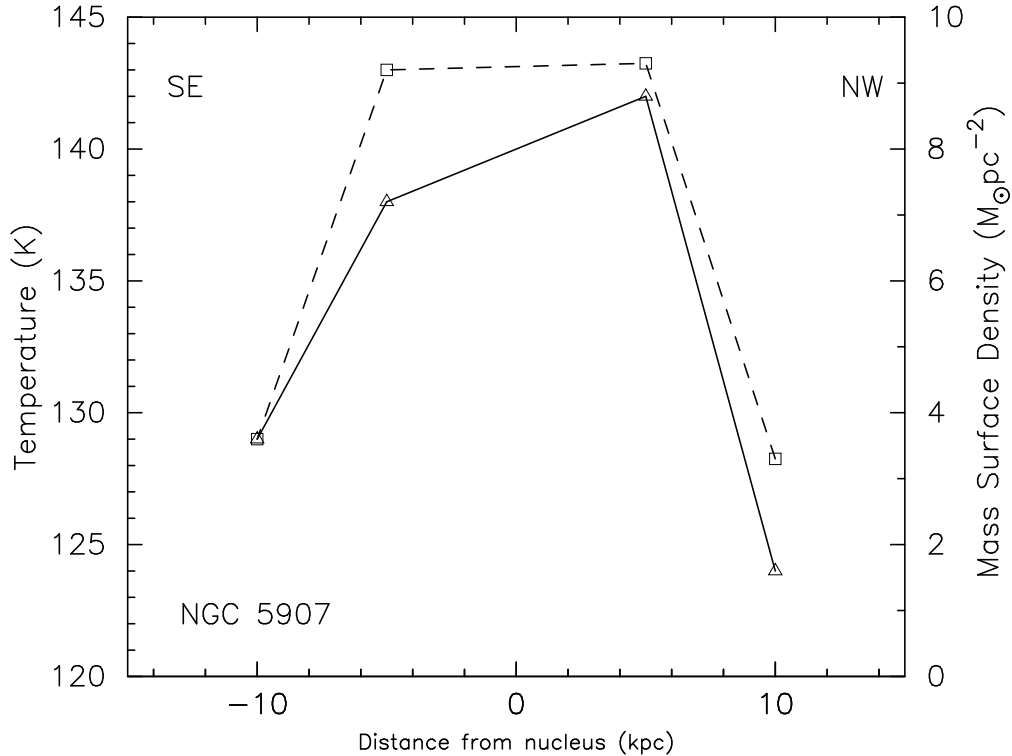}
\caption{Temperature in K (solid line) and the mass surface density in
M$_{\sun}$~pc$^{-2}$ (dashed line) are plotted versus distance from the
nucleus (in kpc) for NGC~5907. The estimates of temperature and mass surface density were derived from the \Ht\ S(0) and S(1) lines. The temperature and 
mass surface density calculations were made with the assumption that the warm molecular \Ht\ gas is uniformly distributed across the slits (see Section~\ref{observ} and \ref{excisection} for discussion on the flux distribution in the slits and the calibration differences between smooth, extended and point source fluxes).}
\label{fig13}
\end{figure*}

The solid lines indicate the best fits to the S(0), S(1), and S(2) data points assuming
\Ht\ in thermal equilibrium with a single-temperature component (we will discuss the
consequences of relaxing this assumption below). The fits also assume a thermal
equilibrium ratio for the ortho-to-para species (O/P), as is reasonable if the density
of the \Ht\ is above the critical density (which, for the low J transitions, is
typically $\sim$~100 mol~cm$^{-3}$), a condition probably satisfied in most cases. For
temperatures less than $\sim$~300 K, this leads to O/P ratios $<$ 3. For example, at T =
115--120 K, O/P = 2 for thermal equilibrium. However, it is far from clear that thermal
equilibrium is appropriate in all cases. For example, if the excitation mechanism were a
shock, then the passage of the shock could leave the \Ht\ molecules in a state where
they do not have enough time to equilibrate. This is another source of uncertainty. For
example, if O/P was 3 instead of 2 (a case where the gas has not had time to come into
thermal equilibrium), then this would change the calculated temperature from T = 120 K
to 113 K with a corresponding increase in the total \Ht\ mass surface density. This
uncertainty in the O/P ratio is comparable with the uncertainty in the clumpiness in the
\Ht\ distribution which leads to the broad range of possible temperatures as shown in
Figures~\ref{fig10} and \ref{fig11}.

The single-temperature fits to these data are shown in each panel of
the figures. Outside the nucleus only an upper limit is available for
the S(2) line. Therefore, fitting more than one thermal component is
not statistically justifiable -- the fits are the formal solutions. We
note that the assumption that the source of \Ht\ is a
point source always yields very low \Ht\ temperatures, bordering on
becoming physically unreasonable. Thus we believe that the warmer
temperatures implied by the extended source calibration are more
physically reasonable for the case of these edge-on galaxies. One
exception is the nucleus of NGC~4565, where a point-source assumption
may be reasonable, as it contains a Seyfert nucleus.

Based on these assumptions, Tables~\ref{Excipars} and \ref{Excipars2} summarize
the derived \Ht\ physical parameters for NGC~4565 and NGC~5907, respectively:
temperature (K), equilibrium ortho/para ratio, column density of \Ht\ (mol
cm$^{-2}$), and the mass surface density of \Ht\ (M$_{\sun}$~pc$^{-2}$). The
temperatures and mass surface densities for NGC~4565 and NGC~5907 are also shown
in Figures~\ref{fig12} and \ref{fig13}. The gas is colder in the outer disk 
where the mass surface density is also lower, creating the apparent impression
of a correlation between temperature and mass surface density. The derived
(extended source) mass surface densities are more than 100 times smaller than
those found in NGC~891 by \citet{vale99}. See Section~\ref{darkmatter} for more 
discussion about the implication of the implied warm \Ht\ mass surface
densities.

If one adopts the extended source assumption, and excludes the nucleus of
NGC~4565 which is significantly warmer, there is no obvious change in the fitted
temperature with radius within the uncertainty from $r$ = 5~kpc to $r$ = 10~kpc
on both sides of this galaxy (T = 146--149 K on the southeastern side and T =
132--134 K on the northwestern side). Even at $r$ = 15~kpc on the northwestern
side, the upper limit to the S(1) flux provides a temperature limit which
is at least consistent with a flat temperature distribution. The situation is
different in NGC~5907, where the outermost 10~kpc points seem more than 10~K
cooler than those measured at 5~kpc. For this galaxy the radial temperature
profile is also symmetric, unlike that in NGC~4565.

In the previous discussion we have made an assumption that a single-temperature
model is reasonable. This is clearly not the case for the nucleus of NGC~4565,
where the 0--0 S(2) line was detected. The first panel of Figure~\ref{fig10}
shows that the single-temperature fits do not pass through the S(2) point.
Indeed, in general, extragalactic sources almost always show a range of
allowable temperatures and often a multiple-component fit is required. One
consequence of fitting a multiple-temperature model is that the warmer component
softens the slope of the fit in the excitation diagram. This means that the
lower temperature component becomes even cooler, once a warm component is
subtracted. To illustrate this we have fitted a two-component model to the
nucleus of NGC~4565 and derived the following temperatures and column densities.
Instead of a single (in this case point-like) nuclear source with T = 122$\pm$4
K and a column density N$_{\rm H_{2}}$ of 5.9~$\times$~10$^{21}$~mol~cm$^{-2}$,
we obtain T(1) =~115$\pm$3 K, N(1)$_{\rm H_{2}}$ =
7.4~$\times$~10$^{21}$~mol~cm$^{-2}$, and T(2) = 450--550 K, N(2)$_{\rm H_{2}}$ =
3--4~$\times$~10$^{18}$~mol~cm$^{-2}$. The warmer component is less constrained
because the error bar on the S(2) line is larger than that of the S(0) and S(1)
lines. Note that the effect in this case of relaxing the single-temperature
model is to increase the cold component column density by 25\%. The warmer
component adds a negligible amount to the final column density.

It is very likely that the nucleus of NGC~4565 is different from the disk because
it contains a Seyfert component which may contribute additional heating to the \Ht\
emission. This is reflected in the generally higher single-temperature fits shown
in Figure~\ref{fig10}. It may seem odd that the two-component fit gives a
temperature for the cold component in the nuclear pointing that is colder than
elsewhere in the disk. However, the thermodynamics of the \Ht\ molecule is likely
to be very complex, involving a multi-phase medium with unknown heating and cooling
conditions. It is also possible that the density distribution of the clouds near
the nucleus is very different from elsewhere in the disk, and there may be more
very dense cold clouds near the nucleus. With the spatial resolution afforded by
the IRS, we cannot resolve this question. Furthermore, magnetohydrodynamic shocks
may be driven into a clumpy medium near the nucleus, which will lead to a range of
temperatures. It should also be noted that it is impossible to estimate what the
separate contributions of the disk and the nucleus are to the observed line fluxes
in the nuclear pointings.

Thus, depending on the strength of a second or third component, the temperature
of the coolest component is always lowered relative to a single-temperature
fit-case. Because we used only single-temperature fits, it is possible that we
underestimated the total \Ht\ column density if warmer components were present,
because a cooler \Ht\ temperature implies a larger total \Ht\ column density.
Since we have, in general, no information about a warmer component, we cannot do
more than fit a single-temperature component and accept a degree of uncertainty
in the final \Ht\ column densities and masses. 

\section{DISCUSSION}

\subsection{Nucleus of NGC~4565}

The line fluxes measured in the Seyfert nucleus of NGC~4565 are listed in
Table~\ref{table1b}. The continuum appears relatively flat, although it shows a
signature of the broad PAH emission feature around 17~$\mu$m. Since the
apertures are relatively large (covering several hundreds of pc), a lot of this
PAH emission is likely to come from the disk of NGC~4565. The 11.3~\mic\ and
12.9~\mic\ PAH features are also strong, but weaker than in the spectra
taken at 5~kpc from the nucleus. The detected emission lines come from \Ht, O,
Ne, S, and Si. The AGN indicator lines of [\ion{Ne}{5}]~14.32~\mic\ and
[\ion{Ne}{5}]~24.31~\mic\ \citep[e.g.,][]{armus04} are both detected at S/N
$\sim$~10. The \Ht\ S(0)/\Ht\ S(1) ratio reaches its minimum at the nuclear
position (see Figure~\ref{fig7}), implying the highest gas temperatures, as can
also be seen in Figure~\ref{fig10}. The [\ion{Ne}{3}]
15.55~\mic/[\ion{Ne}{2}]~12.81~\mic\ ratio reaches a peak in the nucleus. The
value of $\sim$~1 for this ratio indicates a moderate nuclear starburst
\citep{verma02}. It is also consistent with the classification of NGC~4565 as a
Sy1.9 galaxy \citep{deo07}.

\begin{figure}
\centering
\includegraphics[width=8cm]{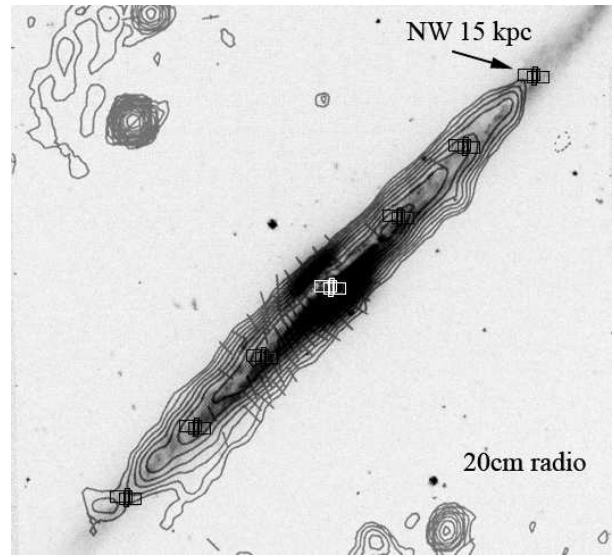}
\caption{20-cm radio continuum emission in NGC~4565 from \citet{suku91}, shown on top of an optical image.}
\label{fig14}
\end{figure}

\subsection{Gas, Dust, and PAH Excitation}
\label{excitdiscussion}

In both galaxies, NGC~4565 and NGC~5907, we see that the emission line
intensities and the derived mass surface densities of \Ht\ emission
(as well as the intensity of the forbidden lines) decrease with
increasing radius, while the temperature decreases only
slightly. Also, the 20-cm radio continuum, for NGC~4565 shown in
Figure~\ref{fig14} \citep{suku91}, decreases strongly towards the
15~kpc radius (which in NGC~4565 is actually outside the detected
radio continuum emission on the NW side).

\begin{figure*}
\centering
\includegraphics[width=10cm,angle=270]{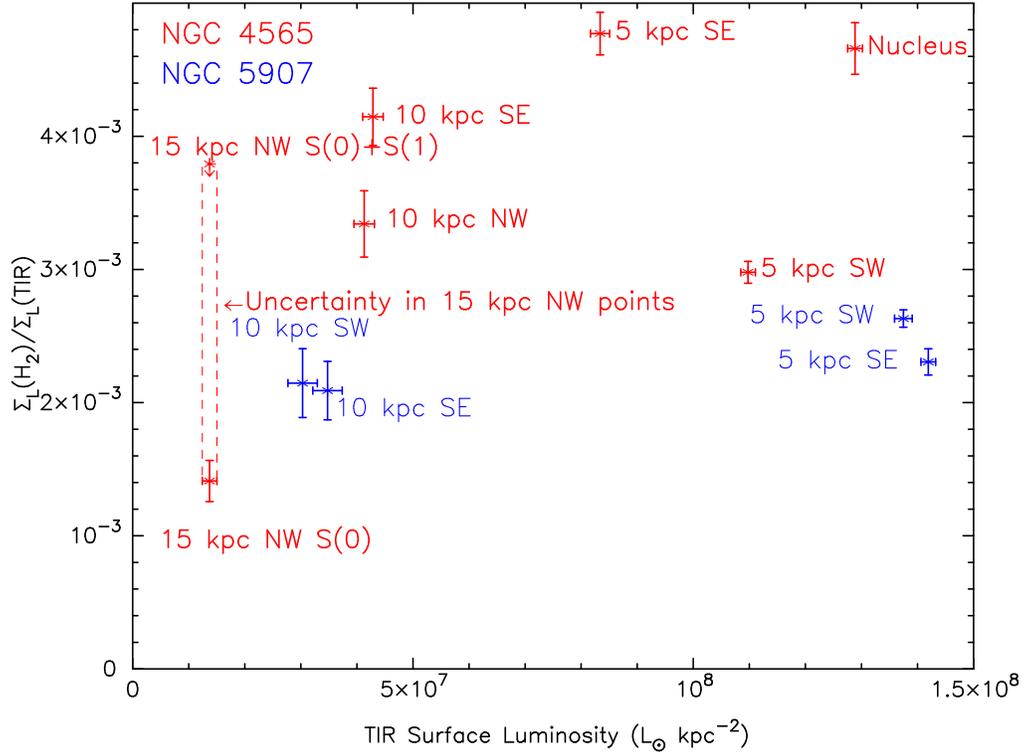}
\caption{Ratios of the \Ht\ and total infrared (TIR) luminosity surface
densities in NGC~4565 and in NGC~5907. Details on the calculation of TIR are given in Section~\ref{excitdiscussion}.}
\label{fig15}
\end{figure*}

\begin{figure*}
\centering
\includegraphics[width=10cm,angle=270]{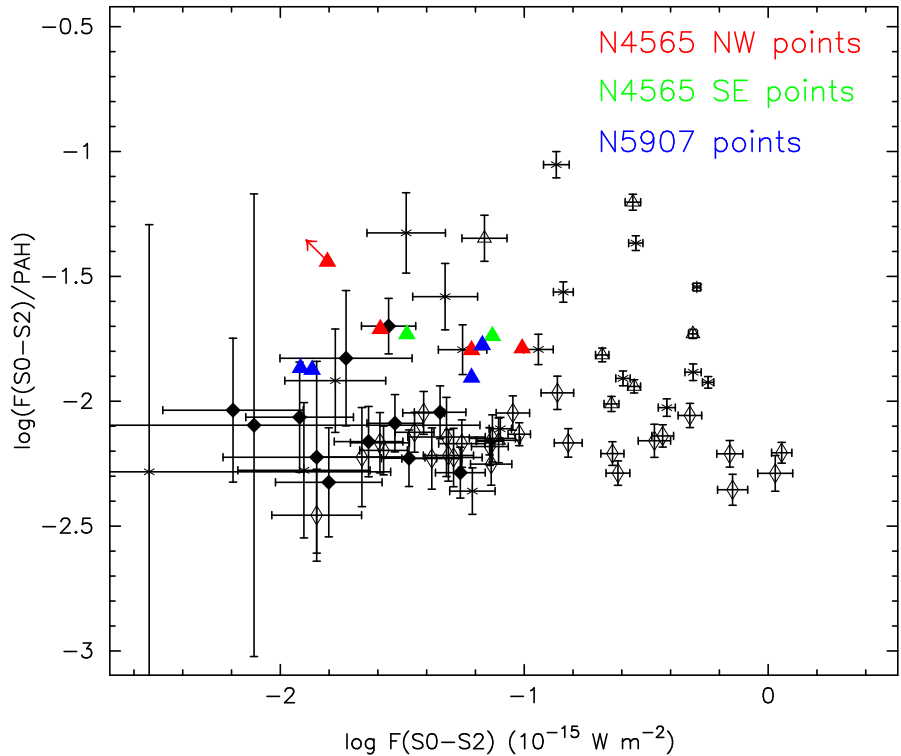}
\caption{Logarithm of the ratios of the \Ht\ emission to the 8~\mic\ emission for NGC~4565 and NGC~5907. NGC~4565 is shown in red and green colors and 
NGC~5907 is shown in blue colors. As in \citet{roussel07}, the open diamonds are for star forming regions, filled diamonds for dwarf galaxies, stars for nuclear regions containing a LINER nucleus, and triangles for nuclear regions containing a Seyfert galaxy.}
\label{fig16}
\end{figure*}

We calculated the ratio of the \Ht\ luminosity surface density over the
total infrared (TIR) emission luminosity surface density 
(Figure~\ref{fig15}). TIR was calculated as in equation (9) of
\citet{bendo08} over the LH slit area, measuring surface brightness
values in the 8, 24, 70, and 160 \mic\ Spitzer IRAC and MIPS maps that
were all smoothed to the resolution of the 160 \mic\ map, at the
positions of the observed IRS slits. This ratio is relatively constant
with the radius at about 0.2\%--0.4\%. This value is somewhat higher
than the 0.05\%--0.1\% typically seen in the SINGS sample, but since
we could not match the resolution and aperture of the broad-band
images, from which TIR was estimated, to the single slit observations
taken in the staring mode, such a bias is expected.

\begin{figure*}
\centering
\includegraphics[width=10cm,angle=270]{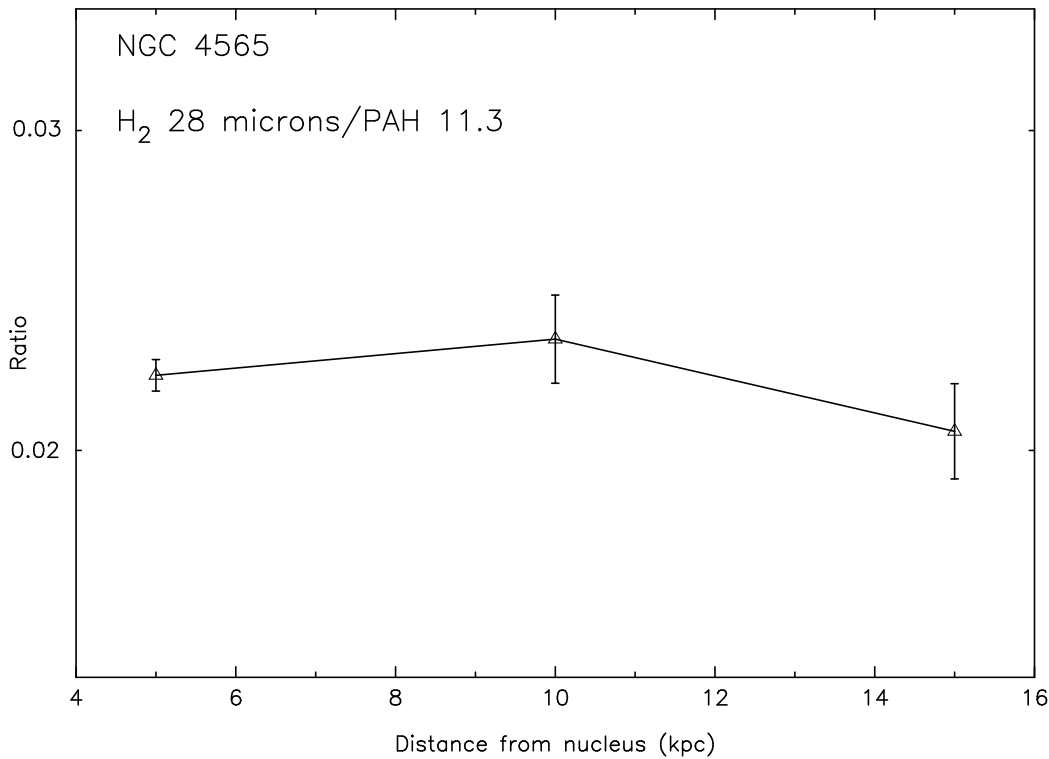}
\caption{Ratios of the fluxes of \Ht\ S(0)~28~\mic\ and the 11.3~\mic\ 
PAH feature are plotted versus the distance from the nucleus (in kpc) 
on the northwestern side of NGC~4565.}
\label{fig17}
\end{figure*}

When plotting the \Ht\ emission power over the IRAC 8~\mic\ power
(Figure~\ref{fig16}) we see that the points in NGC~4565 and NGC~5907
lie generally above the star formation region points in
\citet{roussel07}. Specifically, we see an increase in the ratio
towards the outer 15 kpc NW point in NGC~4565. We also see no change
in the \Ht\ S(0)/11.3~\mic\ PAH ratio (Figure~\ref{fig17}) on the
northwestern side of the disk of NGC~4565, but we see an increase in
the 11.3~\mic/7.7~\mic\ PAH feature ratio (Figure~\ref{fig9}) from
10~kpc to 15~kpc. One explanation is that the PAHs become more neutral
in the lower UV excitation environment of the outer disk at 15~kpc NW
in NGC~4565. The possible change in PAH excitation from ionized to
neutral changes the relative strengths of the 11.3~\mic\ with respect
to the 7.7~\mic\ PAHs because the 11.3~\mic\ PAH feature becomes more
dominant as the PAHs become more neutral. This might naturally
explain why the 15~kpc NW point in Figure~\ref{fig15} stands out. It
is not due to the \Ht\ emission becoming relatively stronger at
15~kpc, but due to the 7.7~\mic\ PAH feature becoming weaker.

We also measured the 24~\mic~flux densities in the areas covered by the IRS
slits in NGC~4565, and noticed that the 24~\mic~flux density decreases with
radius. Since the 24~\mic~emission is a relatively good proxy of the star
formation intensity \citep[e.g.,][]{calzetti07}, this implies that the UV flux
intensity is decreasing with radius, therefore producing a more neutral ISM at
larger radii, consistent with our results derived from the PAH flux ratio. 

\subsection{Warm Molecular Gas Contribution to Dark Mass in the Disks of NGC~4565 and NGC~5907}
\label{darkmatter}

\citet*{hoekstra01} have shown that there is an apparently strong coupling between
the surface densities of neutral hydrogen and dark matter in spiral galaxies, with a
significant and pronounced peak in $\Sigma_{{\rm DM}}$/$\Sigma_{HI}$~$\sim$~9. To see
whether this holds in NGC~5907, we used the multicomponent dynamical model of
NGC~5907 by \citet{barna94}. For ease of calculation we approximated the dark matter
distribution by a singular isothermal sphere. At a radial distance of 10 kpc from the
nucleus we calculate a $\Sigma_{\rm DM}$ = 188~M$_{\sun}$~pc$^{-2}$. $\Sigma_{\rm
HI}$~$\sim$~17.4~M$_{\sun}$~pc$^{-2}$ \citep{sofue94b} and the ratio of the
two is 10.8. Including the warm \Ht\ gas only reduces the dark matter to gas ratio to
9. Thus NGC~5907 appears to follow the relationship found by \citet{hoekstra01}. At a
radial distance of 15~kpc from the nucleus the warm \Ht\ is undetected. Assuming, as suggested by the data, an extended, smooth emission distribution,
$\Sigma_{H_{2}}$~$\le$~3~M$_{\sun}$~pc$^{-2}$. Using our approximation to
the Barnaby--Thronson model, $\Sigma_{\rm DM}$/$\Sigma_{H_{2}}$~$\ge$~42. 
At 15~kpc, $\Sigma_{\rm HI}$ = 10.3~M$_{\sun}$~pc$^{-2}$ and $\Sigma_{\rm DM}$/$\Sigma_{\rm HI}$~= 12.2. 
It is therefore clear that the mass of the ISM in
NGC~5907, including the mass of the warm molecular gas, is too small by more than an
order of magnitude to account for the requisite dark matter. 

Unfortunately there is no model of the mass distribution and dynamics of NGC~4565
similar to that of \citet{barna94} for NGC~5907. The neutral atomic hydrogen 
properties were studied by \citet{rupen91}, and the cold molecular gas properties by
\citet{sofue94a} while visible light surface photometry of NGC~4565 was performed by
\citet{kruit81}. We assumed that the stars and the ISM are confined to a thin disk
and the dark halo can be described, again, by a singular isothermal sphere. The
rotation curve is given by \citet{sofue96,sofue97}. \citet{kruit81} showed that the
optical disk is truncated at a radius of 24.9 kpc, comparable to where
\citet{rupen91} sees a warp. Using a $B$-band luminosity of the old disk of
1.4$\times$10$^{10}$ L$_{\sun}$ and a median value of 7.5 (corrected for Hubble
constant H$_{0}$~=~75~km~s$^{-1}$~Mpc$^{-1}$) for the mass to luminosity ratio of
Sab--Sb galaxies from \citet{roberts94}, the mass of the luminous stellar disk is
10.5$\times$10$^{10}$~M$_{\sun}$. This may be an overestimate as some fraction of the
mass quoted by \citet{roberts94} is dark. The total mass of the neutral atomic
hydrogen is 5.96$\times$10$^{9}$~M$_{\sun}$ \citep{rupen91} and that of the cold
molecular hydrogen 2.4$\times$10$^{9}$~M$_{\sun}$ \citep{sofue94a}. We assumed that
the ISM and the stars are confined to a thin disk and all components are truncated at
25~kpc. \citet{lequeux83} suggests that the velocity at the truncation radius is

\begin{equation}
V^{2} = \frac{GM}{0.6R}
\end{equation}

where $R$ is 25 kpc for NGC~4565 and $M$ is the total mass. Beyond $R$=25 kpc the
velocity is assumed to decline in a Keplerian fashion. At a radial distance of 35~kpc, the observed circular velocity of the galaxy is about 214 km~s$^{-1}$ \citep{sofue97}. From the mass and velocity components of our model we find that the halo contributes a velocity of 150~km~s$^{-1}$ to the system. We reflect these values back to a radius of 15~kpc at which we observed the most distant emission from warm \Ht, and we recalculate the mass surface densities. Assuming a singular isothermal sphere for the dark matter, the mass surface density of dark matter is 86~M$_{\sun}$~pc$^{-2}$, the ratio of the mass surface densities of dark matter and warm 
\Ht\ (with the much likelier smoothly distributed extended source emission calibration) 
is $\ge$~86/1.8 = 48, and that of dark matter to all of the ISM components (neutral 
atomic hydrogen, warm molecular hydrogen, and cold molecular hydrogen) is $\ge$ 15.

From our analysis it is clear that the mass surface densities of the warm molecular gas cannot produce the observed rotation velocities at large radii in NGC~4565 and NGC~5907. The ``missing mass'' 
in these two galaxies cannot be accounted for by warm \Ht\ gas.

\subsection{Source of Excitation of H$_{2}$ in the Outer Disk of NGC~4565}

The molecular gas at the 15 kpc NW point in the outer disk of NGC~4565, as probed by 
the \Ht\ rotational lines, has roughly the same temperature and the same ratio of the
\Ht\ to far-IR power as in the inner disk. However, the star formation rate at the 15
kpc NW point, traced by the mid-IR emission, has substantially decreased, compared to
the inner disk. In other words, although the intensity of the exciting radiation
field has been reduced (as seen also in the change in the ratio of the ionized to neutral PAH molecules and in a reduction in the TIR intensity when comparing the 10 kpc NW and 15
kpc NW points), the molecular gas is heated to a similar temperature throughout the
disk. Therefore, something other than star formation may be heating the gas at
the 15 kpc NW point.

Cosmic ray (CR) heating of the \Ht\ does not appear viable at the 15 kpc NW point
because we see a dramatic decrease in the strength of the synchrotron radio continuum
emission, which is a tracer of cosmic rays accelerated in the magnetic field of the
galaxy, between the 10 and 15 kpc NW points, as shown in Figure~\ref{fig14}. The 15
kpc NW point lies outside the detectable signal in the radio continuum maps of
\citet{suku91}. The upper limit of the 20-cm radio continuum (150 $\mu$Jy/beam;
3$\sigma$) at the 15 kpc NW point suggests a difference of a factor of $>$ 64 in the
20-cm radio continuum flux density between the 10 and 15~kpc points. This change is
not reflected in the decrease in the \Ht\ line luminosity which is only a factor of
7. 

However, despite the lack of detected radio continuum at the 15 kpc NW point, we
cannot completely rule out CR excitation. If we assume an equipartition of energy
between CRs in the disk and the magnetic energy density, the upper limit to the radio
continuum flux density corresponds to an upper limit for the equipartition magnetic
field strength of $B_{{\rm min}}$~$\le$~1~$\mu$G, following the assumptions discussed
in \citet{Govoni04}. This corresponds to a magnetic energy density (and a comparable
CR energy density) of $\sim$~9.6 $\times$ 10$^{-14}$ ergs~cm$^{-3}$. For a canonical
synchrotron lifetime in the mid-plane of 10$^{7}$ yrs, the CRs could potentially
provide $L_{{\rm cr}}$~$\le$ 4.7~$\times$~10$^{30}$~W/kpc$^{2}$ of power if such a
population of CRs existed below the detection limit of the radio continuum
observations. Interestingly, this is only a factor of two lower than the \Ht\
line luminosity in the S(0) line at 15 kpc NW (7 $\times$ 10$^{30}$ W/kpc$^{2}$),
and therefore CR heating, although unlikely (it would require very rapid deposition timescales of much less than 10$^{7}$ yrs and high heating efficiency), cannot be 
completely ruled out. 

We note that the equivalent $B_{{\rm {min}}}$ and $L_{{\rm {cr}}}$ for the 10 kpc NW
point in NGC 4565 are 3.3~$\mu$G and 5.0~$\times$~10$^{-13}$~ergs~cm$^{-3}$,
respectively\footnote{Here we assume that the depth of the radio continuum emitting
region is 10~kpc at $r$~=~10~kpc and 5~kpc at $r$~=~15~kpc.}, and the ratio of
$L_{{\rm {cr}}}$/$L$(\Ht)~$\sim$~0.5 at that point. This suggests that within the
radio continuum emitting disk of NGC~4565, trapped cosmic rays can, in principle,
provide energy to heat the \Ht\ in the disk (again high heating efficiency would be
needed). In those same regions, star formation, through PDR heating, provides a more
likely channel for heating the \Ht\ gas \citep[see][]{roussel07}.

The feasibility of cosmic ray heating at $r$~=~15 kpc can also be estimated using an
independent ionization argument \citep{guillard09a}. If we assume that the cosmic
rays heat the gas through partial ionization of the \Ht\ to H$_{3}$$^{+}$, then the
rate of ionization through cosmic ray heating must balance the rate of cooling of
the \Ht\ per molecule. For a column density $N_{\rm H_{2}}$
$\sim$~1.1~$\times$~10$^{20}$~mol~cm$^{-2}$ and the observed luminosity in the S(0)
line, we estimate the \Ht\ cooling per molecule to be  6.7~$\times$~10$^{-33}$
W~mol$^{-1}$. \citet*{yusef07} estimate the CR heating rate per molecule to be
8~$\times$~10$^{-18}$~$\zeta$$_{\rm H}$ W, where $\zeta$$_{\rm H}$ is the \Ht\
ionization rate. Under these assumptions, for CR heating to balance the \Ht\ cooling
would require an ionization rate of $\sim$~10$^{-15}$~s$^{-1}$. This value is
comparable to that measured in the Galactic Center \citep[see][]{oka05},  but is
unlikely to be realized in the outer disk of NGC~4565. This again suggests  that CR
excitation is an unlikely source of heating for the S(0) line at $r$ = 15 kpc unless
conditions there are very unusual.

The two remaining options for \Ht\ excitation in the outer disk are heating within
extended PDR regions, or shock heating. We have already shown that the ratio of the
\Ht\ power to the far-IR emission power is consistent with PDR heating \citep[in
comparison with the models of][]{draine07}, but it is not clear if this process works
in the outer disk. Indeed, it is very likely (and observations with the {\it Herschel
Space Observatory} will help to resolve this issue) that a large component of the TIR
flux we see in the outer disk of NGC~4565 comes from cirrus clouds heated by the
general radiation field of the outer disk, and not from young stars. Thus one is left
with the puzzling result that the \Ht\ excitation remains constant to within a factor
of two in the outer disk which does not contain a high concentration of young stars.
Widely distributed PDR regions around a smoothly distributed set of faint young
stars may be responsible for the excitation, but this cannot be demonstrated with our observations.

Another possible way of heating the \Ht\ is by shocks, perhaps through a recent
passage of a disturbance through the outer disk of NGC 4565. A recent model of how
\Ht\ can be excited in a powerful shock propagating through a multi-phase medium in
Stephan's Quintet has been presented by \citet{guillard09b}. However, this model was
tuned to the specific problem of how to generate large amounts of power in the \Ht\
lines in a 1000~km~sec$^{-1}$ shock moving through a clumpy medium. To explain the
emission in the outer regions of NGC~4565 via shocks would require considerably less
energy input, but there is no obvious source of energy to drive the shocks.
Curiously, recent {\it Spitzer} observations of a high-latitude cirrus cloud within
the Galaxy \citep{ingalls10} have revealed unusually strong \Ht\
emission from regions that are clearly not associated with PDRs. Shock heating is one
possible explanation. These observations suggest that the excitation of \Ht\ in
galaxy disks is not yet well understood. 

\section{CONCLUSIONS}

We have examined the excitation of gas, dust, and PAHs and the physical conditions of
the ISM in two nearby normal edge-on disk galaxies, NGC~4565 and NGC~5907, out to 15
kpc from the nucleus of each galaxy. Our most important conclusions can be summarized
as follows.

\begin{enumerate}

\item We have detected the rotational 17~\mic\ S(0) and 28~\mic\ S(1) \Ht\ line
transitions at 5 and 10~kpc, and most interestingly, the S(0) \Ht\ line at 15~kpc
NW from the nucleus of NGC~4565.

\item We have discovered that in these two edge-on galaxies, NGC~4565 and
NGC~5907, the warm molecular gas temperature (although uncertain) and the ratio
of the \Ht\ line luminosity surface density to the total infrared luminosity surface
density are rather flat with radius. However, the active star formation rate, as
measured by, e.g., the 24~$\mu$m emission, falls rapidly with radius. This
result is potentially inconsistent with excitation of the \Ht\ emission by photodissociation regions in the outer disks of these galaxies.

\item Alternatives to the \Ht\ excitation in the outer disk are cosmic ray heating and
shocks. Based on the midplane radio continuum emission intensities, excitation by cosmic rays and photodissociation regions are both viable in the inner disk. However, in the outer disk the  non-detection of radio continuum implies that cosmic rays are less important there. Therefore, extended photodissociation regions or shocks can excite the emission at the outermost disk, as seen in NGC~4565 at the 15 kpc NW point.

\item We see an increase of the 11.3~\mic/7.7~\mic\ PAH feature strength ratio (where we
used the IRAC 8~\mic\ band to be a proxy of the 7.7~\mic\ emission) at the 15 kpc NW
position in NGC~4565. We also see that the summed \Ht\ line intensity over the 8~\mic\
emission intensity ratio increases at the same position. Our interpretation is that the
\Ht\ S(0)~28~\mic\ emission at the 15~kpc NW position may still be excited by (weaker)
emission from photodissociation regions, coming from a more neutral medium at this large distance from the nucleus, as the strength of the 7.7~\mic\ PAH feature, which traces more highly ionized dust, decreases with respect to the strength of the 11.3~\mic\ PAH feature, which traces more neutral dust.

\item The observations strongly suggest that the warm molecular gas is smoothly distributed. Assuming such an extended distribution, the detected mass surface densities of warm molecular hydrogen are very low at large radii in both galaxies. It is very unlikely that this component of the ISM contributes at any significant level to the "missing mass" in the outer regions of these two edge-on disk galaxies.

\item The Seyfert 1.9 nucleus in NGC~4565 revealed [\ion{Ne}{5}]~14.32~\mic,
[\ion{Ne}{5}]~24.31~\mic, [\ion{S}{4}]~10.51~\mic, and [\ion{O}{4}]~25.89~\mic\
lines, as well as the 12.28~\mic\ \Ht\ S(2) line. The higher excitation
forbidden lines are expected to be seen in Seyfert nuclei.

\end{enumerate}

\acknowledgments

We are grateful to Tom Jarrett at IPAC for helping us to correct the rogue pixels
with his custom-made IRAF script. We thank the anonymous referee for very helpful and
detailed comments. We acknowledge stimulating discussions with Eric
Murphy on the mid-IR and far-IR properties of nearby galaxies. This research has made
use of the NASA/IPAC Extragalactic Database (NED) which is operated by the Jet
Propulsion Laboratory, California Institute of Technology, under contract with the
National Aeronautics and Space Administration. SMART was developed by the IRS Team at
Cornell University and is available through the {\it Spitzer} Science Center at
Caltech. The IRS was a collaborative venture between Cornell University and Ball
Aerospace Corporation funded by NASA through the Jet Propulsion Laboratory and Ames
Research Center. This work is based on observations made with the {\it Spitzer Space
Telescope}, which is operated by the Jet Propulsion Laboratory, California Institute
of Technology under a contract with NASA. Support for this work was provided by NASA
through an award issued by JPL/Caltech. 

{\it Facilities:} \facility{Spitzer (IRAC, IRS, MIPS)}

\end{document}